\documentclass[a4paper,11pt]{article}
\date{}
\usepackage[latin1]{inputenc}
\usepackage[ruled,vlined,linesnumbered]{algorithm2e}
\usepackage{geometry}
\usepackage{amsmath,amsfonts,enumerate,amssymb,amsthm,color}
\usepackage{fullpage,enumerate}
\usepackage{graphicx}
\usepackage{latexsym,ifthen,epsfig,color}
\usepackage{float,paralist,cite}
\usepackage{version}

\newcommand{\join}{\text{\textcircled{{\footnotesize 1}}}}
\newcommand{\cojoin}{\text{\textcircled{{\footnotesize 0}}}}

\newtheorem{Claim}{Claim}[section]
\newtheorem{theorem}{Theorem}
\newtheorem{lemma}{Lemma}

\newtheorem{corollary}{Corollary}

\newtheorem{proc}{Procedure}[section]
\newtheorem{algo}{Algorithm}[section]

\newcommand{\NP}{\ensuremath{\mathbb{NP}}}

\begin{document}

\author{
Andreas Brandst\"adt\\
\small Institut f\"ur Informatik, Universit\"at Rostock, D-18051 Rostock, Germany\\
\small \texttt{andreas.brandstaedt@uni-rostock.de}\\
\and
Raffaele Mosca\\
\small Dipartimento di Economia, Universit\'a degli Studi ``G. D'Annunzio'',
Pescara 65121, Italy\\
\small \texttt{r.mosca@unich.it}
}

\title{On Efficient Domination for Some Classes of $H$-Free Chordal Graphs}

\maketitle

\begin{abstract}
A vertex set $D$ in a finite undirected graph $G$ is an {\em efficient dominating set} (\emph{e.d.s.}\ for short) of $G$ if every vertex of $G$ is dominated by exactly one vertex of $D$. The \emph{Efficient Domination} (ED) problem, which asks for the existence of an e.d.s.\ in $G$, is known to be \NP-complete even for very restricted graph classes such as for $2P_3$-free chordal graphs while it is solvable in polynomial time for $P_6$-free chordal graphs (and even for $P_6$-free graphs). A standard reduction from the \NP-complete Exact Cover problem shows that ED is \NP-complete for a very special subclass of chordal graphs generalizing split graphs.
The reduction implies that ED is \NP-complete e.g.\ for double-gem-free chordal graphs while it is solvable in linear time for gem-free chordal graphs (by various reasons such as bounded clique-width, distance-hereditary graphs, chordal square etc.), and ED is \NP-complete for butterfly-free chordal graphs while it is solvable in linear time for $2P_2$-free graphs.

We show that (weighted) ED can be solved in polynomial time for $H$-free chordal graphs when $H$ is net, extended gem, or $S_{1,2,3}$.
\end{abstract}

\noindent{\small\textbf{Keywords}:
Weighted efficient domination;
$H$-free chordal graphs;
\NP-completeness;
net-free chordal graphs;
extended-gem-free chordal graphs;
$S_{1,2,3}$-free chordal graphs;
polynomial time algorithm;
clique-width.
}

\section{Introduction}\label{sec:intro}

Let $G=(V,E)$ be a finite undirected graph. A vertex $v$ {\em dominates} itself and its neighbors. A vertex subset $D \subseteq V$ is an {\em efficient dominating set} ({\em e.d.s.}\ for short) of $G$ if every vertex of $G$ is dominated by exactly one vertex in $D$; for any e.d.s.\ $D$ of $G$, $|D \cap N[v]| = 1$ for every $v \in V$ (where $N[v]$ denotes the closed neighborhood of $v$).
Note that not every graph has an e.d.s.; the {\sc Efficient Dominating Set} (ED) problem asks for the existence of an e.d.s.\ in a given graph~$G$.

\medskip

The {\sc Exact Cover} problem asks for a subset ${\cal F'}$ of a set family ${\cal F}$ over a ground set, say $V$, containing every vertex in $V$ exactly once. In particular, this means that the elements of ${\cal F'}$ form a partition of $V$, i.e., for every two distinct elements $U,W \in {\cal F'}$, $U \cap W = \emptyset$ and $\bigcup_{X \in {\cal F'}}=V$. Thus, {\sc Exact Cover} is a partition problem since it asks for a subset ${\cal F'}$ of ${\cal F}$ which forms a partition of $V$ (however, in \cite{GarJoh1979}, the problem {\sc Partition} is a distinct problem [SP12]).
As shown by Karp \cite{Karp1972}, {\sc Exact Cover} is \NP-complete even for set families containing only $3$-element subsets of $V$ (see problem X3C [SP2] in \cite{GarJoh1979}).

Clearly, ED is {\sc Exact Cover} for the closed neighborhood hypergraph of $G$.
The notion of efficient domination was introduced by Biggs \cite{Biggs1973} under the name {\em perfect code}.
The ED problem is motivated by various applications, including coding theory and resource allocation in parallel computer networks; see e.g.\
\cite{BanBarSla1988,BanBarHosSla1996,Biggs1973,ChaPanCoo1995,LiaLuTan1997,Lin1998,LivSto1988,LuTan2002,Milan2012,Yen1992,YenLee1996}.

\medskip

In \cite{BanBarSla1988,BanBarHosSla1996}, it was shown that the ED problem is \NP-complete. Moreover, ED is \NP-complete for $2P_3$-free chordal unipolar graphs~\cite{EscWan2014,SmaSla1995,YenLee1996}.

\medskip

In this paper, we will also consider the following weighted version of the ED problem:
\medskip
\begin{center}
\fbox{\parbox{0.82\linewidth}{\noindent
{\sc Weighted Efficient Domination (WED)}\\[.8ex]
\begin{tabular*}{.9\textwidth}{rl}
{\em Instance:} & A graph $G=(V,E)$, vertex weights $\omega:V\to \mathbb{N} \cup \{\infty\}$.\\
{\em Task:} & Find an e.d.s.\ of minimum finite total weight,\\
&  or determine that $G$ contains no such e.d.s.
\end{tabular*}
}}
\end{center}

The relationship between WED and ED is analyzed in \cite{BraFicLeiMil2013}.

\medskip

For a set ${\cal F}$ of graphs, a graph $G$ is called {\em ${\cal F}$-free} if $G$ contains no induced subgraph isomorphic to a member of ${\cal F}$.
In particular, we say that $G$ is {\em $H$-free} if $G$ is $\{H\}$-free.
Let $H_1+H_2$ denote the disjoint union of graphs $H_1$ and $H_2$, and for $k \ge 2$, let $kH$ denote the disjoint union of $k$ copies of $H$.
For $i \ge 1$, let $P_i$ denote the chordless path with $i$ vertices, and let $K_i$ denote the complete graph with $i$ vertices (clearly, $P_i=K_i$ for $i=1,2$).
For $i \ge 4$, let $C_i$ denote the chordless cycle with $i$ vertices.

For indices $i,j,k \ge 0$, let $S_{i,j,k}$ denote the graph with vertices $u,x_1,\ldots,x_i$, $y_1,\ldots,y_j$, $z_1,\ldots,z_k$ such that the subgraph induced by $u,x_1,\ldots,x_i$ forms a $P_{i+1}$ $(u,x_1,\ldots,x_i)$, the subgraph induced by $u,y_1,\ldots,y_j$ forms a $P_{j+1}$ $(u,y_1,\ldots,y_j)$, and the subgraph induced by $u,z_1,\ldots,z_k$ forms a $P_{k+1}$ $(u,z_1,\ldots,z_k)$, and there are no other edges in $S_{i,j,k}$. Thus, claw is $S_{1,1,1}$, chair is $S_{1,1,2}$, and $P_k$ is isomorphic to $S_{0,0,k-1}$. Claw will also be denoted by $K_{1,3}$, and its {\em midpoint} is the vertex with degree 3 in the claw.

$H$ is a {\em linear forest} if every component of $H$ is a chordless path, i.e., $H$ is claw-free and cycle-free.

$H$ is a {\em co-chair} if it is the complement graph of a chair. $H$ is a $P$ if $H$ has five vertices such that four of them induce a $C_4$ and the fifth is adjacent to exactly one of the $C_4$-vertices. $H$ is a co-$P$ if $H$ is the complement graph of a $P$. $H$ is a {\em bull} if $H$ has five vertices such that four of them induce a $P_4$ and the fifth is adjacent to exactly the two mid-points of the $P_4$. $H$ is a {\em net} if $H$ has six vertices such that five of them induce a bull and the sixth is adjacent to exactly the vertex of the bull with degree 2. $H$ is a {\em diamond} if $H$ has four vertices such that only two of them are nonadjacent. The diamond will also be denoted by $K_4-e$.
$H$ is a {\em gem} if $H$ has five vertices such that four of them induce a $P_4$ and the fifth is adjacent to all of the $P_4$ vertices. $H$ is a {\em co-gem} if $H$ is the complement graph of a gem.

\medskip

For a vertex $v \in V$, $N(v)=\{u \in V: uv \in E\}$ denotes its ({\em open}) {\em neighborhood}, and $N[v]=\{v\} \cup N(v)$ denotes its {\em closed neighborhood}. A vertex $v$ {\em sees} the vertices in $N(v)$ and {\em misses} all the others.
The {\em non-neighborhood} of a vertex $v$ is $\overline{N}(v):=V \setminus N[v]$.
For $U \subseteq V$, $N(U):= \bigcup_{u \in U} N(u) \setminus U$ and $\overline{N}(U):=V \setminus(U \cup N(U))$.

We say that for a vertex set $X\subseteq V$, a vertex $v \notin X$ has a join (resp.,~co-join) to $X$ if $X\subseteq N(v)$ (resp., $X\subseteq \overline{N}(v)$).
Join (resp., co-join) of $v$ to $X$ is denoted by $v \join X$ (resp., $v \cojoin X$). Correspondingly, for vertex sets $X,Y \subseteq V$ with $X \cap Y = \emptyset$,
$X \join Y$ denotes $x \join Y$ for all $x \in X$ and $X \cojoin Y$ denotes $x \cojoin Y$ for all $x \in X$. A vertex $x \notin U$ {\em contacts $U$} if $x$ has a neighbor in $U$. For vertex sets $U,U'$ with $U \cap U' = \emptyset$, $U$ {\em contacts $U'$} if there is a vertex in $U$ contacting $U'$.

If $v\not\in X$ but $v$ has neither a join nor a co-join to $X$, then we say that $v$ {\it distinguishes} $X$.
A set $H$ of at least two vertices of a graph $G$ is called \emph{homogeneous} if $H \not= V(G)$ and every vertex outside $H$ is either adjacent to all vertices in $H$, or to no vertex in $H$.
Obviously, $H$ is homogeneous in $G$ if and only if $H$ is homogeneous in the complement graph $\overline{G}$.
A graph is {\em prime} if it contains no homogeneous set. In \cite{BraGia2014,BraMilNev2013}, it is shown that the WED problem can be reduced to prime graphs.

\medskip

A graph $G$ is {\em chordal} if it is $C_i$-free for any $i \ge 4$. $G=(V,E)$ is {\em unipolar} if $V$ can be partitioned into a clique and the disjoint union of cliques, i.e., there is a partition $V=A \cup B$ such that $G[A]$ is a complete subgraph and $G[B]$ is $P_3$-free. $G$ is a {\em split graph} if $G$ and its complement graph are chordal. Equivalently, $G$ can be partitioned into a clique and an independent set. It is well known that $G$ is a split graph if and only if it is ($2P_2,C_4,C_5$)-free \cite{FoeHam1977}.

\medskip

It is well known that ED is \NP-complete for claw-free graphs (even for ($K_{1,3},K_4-e$)-free perfect graphs \cite{LuTan1998}) as well as for bipartite graphs (and thus for triangle-free graphs) \cite{LuTan2002} and for chordal graphs \cite{EscWan2014,SmaSla1995,YenLee1996}. Thus, for the complexity of ED on $H$-free graphs,
the most interesting cases are when $H$ is a linear forest. Since (W)ED is \NP-complete for $2P_3$-free graphs and polynomial for $(P_5+kP_2)$-free graphs \cite{BraGia2014,BraGiaMil2018}, the class of $P_6$-free graphs was the only open case. It was finally solved in \cite{BraMos2015,BraMos2016} by a direct polynomial time approach (and in \cite{LokPilvan2015} by an indirect one).

\medskip

It is well known that for a graph class with bounded clique-width, ED can be solved in polynomial time \cite{CouMakRot2000}. Thus we only consider ED on $H$-free chordal graphs for which the clique-width is unbounded. For example, the clique-width of $H$-free chordal graphs is unbounded for claw-free chordal graphs while it is bounded if $H \in\{$bull, gem, co-gem, co-chair$\}$. In \cite{BraDabHuaPau2015}, the clique-width of $H$-free chordal graphs is classified for all but two stubborn cases.

\medskip

For graph $G=(V,E)$, let $d_G(x,y)$ denote the distance between $x$ and $y$ (i.e., the shortest length of a path between $x$ and $y$) in $G$.
The square $G^2$ has the same vertex set $V$ as $G$, and two vertices $x,y \in V$, $x \neq y$, are adjacent in $G^2$ if and only if $d_G(x,y) \le 2$.
The WED problem on $G$ can be reduced to Maximum Weight Independent Set (MWIS) on $G^2$ (see \cite{BraLeiRau2012,BraFicLeiMil2013,BraMilNev2013,Milan2012}).

While the complexity of ED for $2P_3$-free chordal graphs is \NP-complete (as mentioned above), it was shown in \cite{BraEscFri2015} that WED is solvable in polynomial time for $P_6$-free chordal graphs, since the square of every $P_6$-free chordal graph $G$ with e.d.s.\ is also chordal.

It is well known \cite{Frank1975} that MWIS is solvable in linear time for chordal graphs.

\medskip

However, there are still many cases of graphs $H$ for which the complexity of WED in $H$-free chordal graphs is open.

\section{WED is \NP-Complete for Chordal Hereditary Satgraphs}\label{WEDNPchordalsatgr}

It is well known \cite{ChaLiu1993} that WED is solvable in linear time for split graphs. In this section, we show that ED is \NP-complete for a slight generalization of split graphs, namely a subclass of chordal hereditary satgraphs: A graph $G$ is called a {\em satgraph} (described by Zverovich in \cite{Zvero2006}) if there exists a partition $A \cup B =V(G)$ such that

\begin{itemize}
\item[$(i)$] $A$ induces a complete subgraph (possibly, $A=\emptyset$),
\item[$(ii)$] $G[B]$ is an induced matching (possibly, $B=\emptyset$), and
\item[$(iii)$] there are no triangles $(a,b,b')$, where $a \in A$ and $b,b' \in B$.
\end{itemize}

In  \cite{Zvero2006}, Zverovich characterized the class of hereditary satgraphs as the class of ${\cal Z}_{SAT}$-free graphs where the set ${\cal Z}_{SAT}$ consists of the graphs $F_1,F_2,\ldots,F_{21}$ shown in Figure 3 of \cite{Zvero2006}. Hereby, $F_i$ for $i \in \{1,2,4,7,8,13,14,15,16,18,19,20,21\}$ contain $C_4,C_5,C_6$ or $C_7$.

\begin{figure}[ht]
 \begin{center}
 \epsfig{file=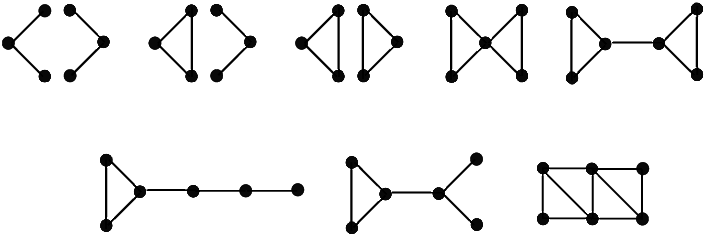}
    \caption{$2P_3$, $K_3+P_3$, $2K_3$, butterfly, extended butterfly, extended co-$P$, extended chair, and double-gem}
   \label{butterfly}
  \end{center}
\end{figure}

The eight remaining $F_i$, namely $F_3,F_5,F_6,F_9,F_{10},F_{11},F_{12},F_{17}$ are presented in Figure \ref{butterfly}.

\begin{lemma}\label{butterflyNPc}
ED is \NP-complete for $(2P_3$, $K_3+P_3$, $2K_3$, butterfly, extended butterfly, extended co-$P$, extended chair, double-gem$)$-free chordal and unipolar graphs.
\end{lemma}

\noindent
{\bf Proof.} The reduction from X3C to Efficient Domination will show that ED is \NP-complete for this special subclass of chordal graphs.

Let $H=(V,{\cal E})$ with $V=\{v_1,\ldots,v_n\}$ and ${\cal E}=\{e_1,\ldots,e_m\}$ be a hypergraph with $|e_i|=3$ for all $i \in \{1,\ldots,m\}$. Let $G_H$ be the following reduction graph:

$V(G_H)=V \cup X \cup Y$ such that $X=\{x_1,\ldots,x_m\}$, $Y=\{y_1,\ldots,y_m\}$ and $V,X,Y$ are pairwise disjoint. The edge set of $G_H$ consists of all edges
$v_ix_j$ whenever $v_i \in e_j$. Moreover $V$ is a clique in $G_H$, $X$ is an independent subset in $G_H$, and every $y_i$, $i=1,\ldots,m$, is only adjacent to $x_i$.

Clearly, $H=(V,{\cal E})$ has an exact cover if and only if $G_H$ has an e.d.s.\ $D$: For an exact cover ${\cal E}'$ of $H$, every $e_i \in {\cal E}'$ corresponds to vertex $x_i \in D$, and every $e_i \notin {\cal E}'$ corresponds to vertex $y_i \in D$.
Conversely, let $D$ be an e.d.s.\ in $G_H$. If $D \cap V \neq \emptyset$, say without loss of generality, $v_1 \in V \cap D$ and $v_1 \in e_1$ then $v_1$ dominates $x_1$ and $y_1$ cannot be dominated which is a contradiction.
Thus, we have $D \cap V=\emptyset$, and now, $D \cap X$ corresponds to an exact cover of $H$.

\medskip

Clearly, $G_H$ is chordal and unipolar. Since any induced $P_3$ or $K_3$ in $G_H$ has a vertex in $V$, the reduction shows that $G_H$ is not only $2P_3$-free but also $F$-free for various other graphs $F$ such as $K_3+P_3$, $2K_3$, butterfly, extended butterfly, extended co-$P$, extended chair, and double-gem as shown in Figure \ref{butterfly}.
\qed

\medskip

The reduction implies that WED is \NP-complete e.g.\ for double-gem-free chordal graphs while it is solvable in linear time for gem-free chordal graphs (since
gem-free chordal graphs are distance-hereditary and thus, their clique-width is at most 3 as shown in \cite{GolRot2000}), and WED is \NP-complete for butterfly-free chordal graphs while it is solvable in linear time for $2P_2$-free graphs \cite{BraMilNev2013}.

\begin{figure}[ht]
  \begin{center}
   \epsfig{file=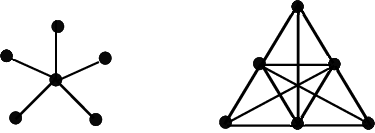}
   \caption{$K_{1,5}$ and $K_3 \join 3K_1$}
   \label{K15etc}
  \end{center}
\end{figure}

\begin{lemma}\label{EDK15frNPc}
ED is \NP-complete for $K_{1,5}$-free chordal graphs and for $K_3 \join 3K_1$-free chordal graphs.
\end{lemma}

\noindent
{\bf Proof.} The Exact Cover problem remains \NP-complete if no element occurs in more than three subsets (see X3C [SP2] in \cite{GarJoh1979}).
With respect to the standard reduction, recall that $V(G_H)=V \cup X \cup Y$, $V$ is a clique in $G_H$, for each hyperedge $e_i \in {\cal E}$, there is exactly one vertex $x_i \in X$ that corresponds to $e_i$, $X$ is independent in $G_H$, and for every $y_i \in Y$, $x_i$ is the only neighbor of $y_i$ in $G_H$.

We first claim that every midpoint of a claw in $G_H$ is in $V$: Let $a,b,c,d$ induce a claw in $G_H$ with midpoint $a$. Then obviously, $a \notin Y$, at most one of $b,c,d$ is in $V$, and if $a \notin V$, i.e., $a \in X$ then two of $b,c,d$ are in $V$ which is a contradiction.

Now $G_H$ is $K_{1,5}$-free since for $K_{1,5}$, say with vertices $a,b,c,d,e,f$ and midpoint $a$, we have $a \in V$ and at most one of $b,c,d,e,f$ is in $V$, say $b \in V$ but then $c,d,e,f \in X$ which is a contradiction to the Exact Cover condition that no element occurs in more than three subsets.

Finally, we claim that $G_H$ is $K_3 \join 3K_1$-free: Let $a,b,c,d,e,f$ induce a $K_3 \join 3K_1$ such that $a,b,c$ induce a $K_3$ and $d,e,f$ induce a $3K_1$. Then each of $a,b,c$ are midpoint of a claw, and thus, $a,b,c \in V$. Moreover, at most one of $d,e,f$ is in $V$, say $e,f \in X$ but now, $e$ and $f$ have a join to the same hyperedge $\{a,b,c\}$ which is a contradiction to the standard reduction.
\qed

\section{$G^2$-Approach For Net-Free and Extended-Gem-Free Chordal Graphs}\label{WEDnetextgemfreechordal}

\begin{figure}[ht]
  \begin{center}
   \epsfig{file=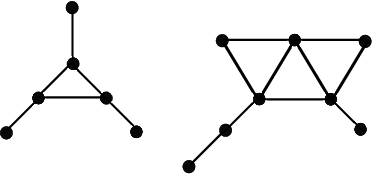}
   \caption{net and extended gem}
   \label{netextgem}
  \end{center}
\end{figure}

Motivated by the $G^2$ approach in \cite{BraEscFri2015,BraEscFriKar2017}, and the result of Milani\v c \cite{Milan2012} showing that for ($S_{1,2,2}$,net)-free graphs $G$, its square
 $G^2$ is claw-free, we show in this section that $G^2$ is chordal for $H$-free chordal graphs with e.d.s.\ when $H$ is a net or an extended gem (see Figure \ref{netextgem} - extended gem generalizes $S_{1,2,2}$ and some other subgraphs), and thus, WED is solvable in polynomial time for these two graph classes.

\begin{Claim}\label{CkinG2}
Let $G$ be a chordal graph, and let $v_1,\ldots,v_k$, $k \ge 4$, induce a $C_k$ in $G^2$ with $d_G(v_i,v_{i+1}) \le 2$ and $d_G(v_i,v_j) \ge 3$, $i,j \in \{1,\ldots,k\}$, $|i-j|>1$ $($index arithmetic modulo $k)$. Then we have:
\begin{itemize}
\item[$(i)$] For each $i \in \{1,\ldots,k\}$, $d_G(v_i,v_{i+1}) = 2$.
\item[$(ii)$] Let $x_i$ be a common neighbor of $v_i$ and $v_{i+1}$ in $G$ $($an {\em auxiliary vertex}$)$. Then for each $i,j \in \{1,\ldots,k\}$, $i \neq j$, we have $x_i \neq x_j$, and $x_ix_{i+1} \in E(G)$.
\end{itemize}
\end{Claim}

\noindent
{\em Proof.}
(i): Suppose without loss of generality that $d_G(v_1,v_2)=1$. Then, since $d_G(v_1,v_3) \ge 3$ and $d_G(v_k,v_2) \ge 3$, we have
 $d_G(v_2,v_3)=2$ and $d_G(v_k,v_1)=2$; let $x_2$ be a common neighbor of $v_2,v_3$ and $x_k$ be a common neighbor of $v_k,v_1$. Clearly, $x_2 \neq x_k$ since $d_G(v_k,v_2) \ge 3$. Moreover, $x_2v_1 \notin E$ since $d_G(v_1,v_3) \ge 3$ and $x_kv_2 \notin E$ since $d_G(v_k,v_2) \ge 3$. Now, $x_kx_2 \notin E$ since otherwise $x_k,v_1,v_2,x_2$ would induce a $C_4$ in $G$ but now in any case, the $P_4$ induced by $x_k,v_1,v_2,x_2$ leads to a chordless cycle in $G$ which is a contradiction.

\noindent
(ii): Clearly, as above, we have $x_i \neq x_j$ for any $i \neq j$. Without loss of generality, suppose to the contrary that there is a non-edge $x_kx_1 \notin E$.
Then, if $x_k$ and $x_1$ have a common neighbor $x_i$, $i \neq k,1$, then $x_k,v_1,x_1,x_i$ would induce a $C_4$ in $G$ which is a contradiction, and if
$x_k$ and $x_1$ do not have have any common neighbor $x_i$, $i \neq k,1$, then a shortest path between $x_1$ and $x_k$ in $G[\{x_1,v_2,x_2,v_3,\ldots,x_{k-1},v_k,x_k\}]$ together with $v_1$ would again lead to a chordless cycle in $G$ which is a contradiction.
\qed

\begin{theorem}\label{EDnetfrchordalG2}
If $G$ is a net-free chordal graph with e.d.s.\ then $G^2$ is chordal.
\end{theorem}

\noindent
{\bf Proof.} Let $G=(V,E)$ be a net-free chordal graph and assume that $G$ has an e.d.s.\ $D$. We first show that $G^2$ is $C_4$-free:

\medskip

Suppose to the contrary that $G^2$ contains a $C_4$, say with vertices $v_1,v_2,v_3,v_4$ such that $d_G(v_i,v_{i+1}) \le 2$ and $d_G(v_i,v_{i+2}) \ge 3$, $i \in \{1,2,3,4\}$ (index arithmetic modulo 4). By Claim~\ref{CkinG2}, we have $d_G(v_i,v_{i+1}) = 2$ for each $i \in \{1,2,3,4\}$; let $x_i$ be a common neighbor of $v_i,v_{i+1}$. By Claim~\ref{CkinG2}, $x_i \neq x_j$ for $i \neq j$. Since $G$ is chordal, $x_1,x_2,x_3,x_4$ either induce a diamond or $K_4$ in $G$.

Assume first that $x_1,x_2,x_3,x_4$ induce a diamond in $G$, say with $x_1x_3 \in E$ and $x_2x_4 \notin E$.
We claim:
\begin{equation}\label{diamondxinotinDnetfr}
D \cap \{x_1,x_2,x_3,x_4\} = \emptyset.
\end{equation}

\noindent
{\em Proof.}
First suppose to the contrary that $x_1 \in D$. Then by the e.d.s.\ property, we have $v_3,v_4,x_2,x_3,x_4 \notin D$. Since $v_3$ and $v_4$ have to be dominated by $D$, let $d_3 \in D$ with $d_3v_3 \in E$ and $d_4 \in D$ with $d_4v_4 \in E$. Clearly, $d_3 \neq x_2,x_3$ and $d_4 \neq x_3,x_4$.
By the e.d.s.\ property, $d_3$ and $d_4$ are nonadjacent to the neighbors $v_1,v_2,x_2,x_3,x_4$ of $x_1$. Thus, $d_3 \neq d_4$ since otherwise $x_1,x_2,v_3,d_3,v_4,x_4$ would induce a $C_6$ in the chordal graph $G$. This implies $d_3v_4 \notin E$ but now, $v_2,x_2,v_3,d_3,x_3,v_4$ induce a net in $G$ which is a contradiction. Thus, $x_1 \notin D$ and correspondingly, $x_3 \notin D$.

Now suppose to the contrary that $x_2 \in D$. Then by the e.d.s.\ property, $v_1,v_4,x_1,x_3,x_4 \notin D$. Since $v_1$ and $v_4$ have to be dominated by $D$,
let $d_1 \in D$ with $d_1v_1 \in E$ and $d_4 \in D$ with $d_4v_4 \in E$. Clearly, $d_1 \neq x_1,x_4$ and $d_4 \neq x_3,x_4$. By the e.d.s.\ property, $d_1$ and $d_4$ are nonadjacent to the neighbors $v_2,v_3,x_1,x_3$ of $x_2$. Thus, $d_1v_4 \notin E$ since otherwise $d_1,v_1,x_1,x_3,v_4$ would induce a $C_5$ in the chordal graph $G$, and analogously, $d_4v_1 \notin E$.
Now, if $d_1x_4 \notin E$ then $d_1,v_1,x_1,v_2,x_4,v_4$ induce a net in $G$, and if $d_1x_4 \in E$ then by the e.d.s.\ property, $d_4x_4 \notin E$ and thus,
$d_4,v_4,x_3,v_3,x_4,v_1$ induce a net in $G$, which is a contradiction.
Thus, $x_2 \notin D$ and correspondingly, $x_4 \notin D$, and claim (\ref{diamondxinotinDnetfr}) is shown.
$\diamond$

\medskip

Next we claim:
\begin{equation}\label{diamondvinotinDnetfr}
D \cap \{v_1,v_2,v_3,v_4\} = \emptyset.
\end{equation}

\noindent
{\em Proof.}
Without loss of generality, suppose to the contrary that $v_1 \in D$. Then by the e.d.s.\ property, we have $v_2,v_4,x_1,x_2,x_3,x_4 \notin D$. Since $v_2$ and $v_4$ have to be dominated by $D$, let $d_2 \in D$ with $d_2v_2 \in E$ and $d_4 \in D$ with $d_4v_4 \in E$.
Since $d_G(v_2,v_4) > 2$, we have $d_2 \neq d_4$.

Moreover, $d_2x_3 \notin E$ since otherwise, $d_2,v_2,x_1,x_3$ induce a $C_4$ in $G$. This implies $d_2v_3 \notin E$ since otherwise,
$d_2,v_3,x_3,x_1,v_2$ induce a $C_5$ in $G$.

Now, if $d_2x_2 \notin E$ then $d_2,v_2,x_2,v_3,x_1,v_1$ induce a net, and if $d_2x_2 \in E$ then $d_2,x_2,x_1,x_3,v_1,v_4$ induce a net, which is a contradiction.

Thus, $v_1 \notin D$, and correspondingly, $v_2,v_3,v_4 \notin D$, and claim (\ref{diamondvinotinDnetfr}) is shown.
$\diamond$

\medskip

Let $d_i \in D$ be the $D$-neighbor of $v_i$. By (\ref{diamondxinotinDnetfr}) and (\ref{diamondvinotinDnetfr}) and the distance properties, we have $d_i \neq v_j,x_j$, $i,j \in \{1,2,3,4\}$.
Next we claim that $d_1,d_2,d_3,d_4$ are pairwise distinct:
\begin{equation}\label{diamondd1d2d3d4distinctinDnetfr}
|\{d_1,d_2,d_3,d_4\}|=4.
\end{equation}

\noindent
{\em Proof.} Since $d_G(v_1,v_3) > 2$ and $d_G(v_2,v_4) > 2$, we have $d_1 \neq d_3$ and $d_2 \neq d_4$. Thus, $|\{d_1,d_2,d_3,d_4\}| \ge 2$.

If without loss of generality, $d_1=d_4$, i.e., $d_1v_1 \in E$ and $d_1v_4 \in E$ then, since $d_1,v_1,x_1,x_3,v_4$ do not induce a $C_5$ in $G$, we have $d_1x_1 \in E$ or $d_1x_3 \in E$, and if without loss of generality, $d_1x_1 \in E$ and $d_1x_3 \notin E$ then $d_1,x_1,x_3,v_4$ induce a $C_4$ in $G$. Thus, $d_1x_1 \in E$ and $d_1x_3 \in E$.

This shows that if $d_1v_1 \in E$ and $d_1v_4 \in E$ then $d_2 \neq d_3$, and thus $|\{d_1,d_2,d_3,d_4\}| \ge 3$.

Now assume that $|\{d_1,d_2,d_3,d_4\}| = 3$, i.e., $d_1v_1 \in E$ and $d_1v_4 \in E$, $d_2v_3 \notin E$ and $d_3v_2 \notin E$.
Recall $d_1x_1 \in E$ and $d_1x_3 \in E$. Thus, $d_2x_1 \notin E$, $d_2x_3 \notin E$, $d_3x_1 \notin E$, $d_3x_3 \notin E$.

If $d_2x_2 \notin E$ then $d_2,v_2,x_1,v_1,x_2,v_3$ induce a net in $G$, and if $d_2x_2 \in E$ then $d_3x_2 \notin E$ and thus, $d_3,v_3,x_2,v_2,x_3,v_4$ induce a net in $G$ which is a contradiction. Thus, $d_1,d_2,d_3,d_4$ are pairwise distinct, and claim (\ref{diamondd1d2d3d4distinctinDnetfr}) is shown.
$\diamond$

\medskip

If $d_1x_1 \notin E$ and $d_1x_4 \notin E$ then $d_1,v_1,x_1,x_4,v_2,v_4$ induce a net in $G$, and correspondingly by symmetry, a similar statement can be made about $d_i,x_{i-1},x_i$, $i \neq 1$. Thus, we can assume that for each $i \in \{1,\ldots,4\}$, $d_i$ sees at least one of $x_{i-1},x_i$ (index arithmetic modulo $4$).

If $d_1x_1 \in E$ and $d_1x_4 \in E$ then clearly, $d_2x_1 \notin E$ and $d_4x_4 \notin E$ and thus, by the above, we can assume that $d_2x_2 \in E$ and
$d_4x_3 \in E$ but now, $d_2,x_2,v_3,x_3,d_3,d_4$ induce a net in $G$.

Thus, assume that $d_1$ is adjacent to exactly one of $x_1,x_4$, say $d_1x_1 \in E$ (which implies $d_2x_1 \notin E$) and $d_1x_4 \notin E$.
By symmetry, this holds for $d_2,d_3,d_4$ as well, i.e., $d_2x_2 \in E$, $d_3x_3 \in E$, and $d_4x_4 \in E$. Then $d_1,x_1,d_2,x_2,d_3,x_3$ induce a net in $G$.

\medskip

Thus, when $x_1,x_2,x_3,x_4$ induce a diamond in $G$, then $G^2$ does not contain a $C_4$ with vertices $v_1,v_2,v_3,v_4$.

\medskip

Now assume that $x_1,x_2,x_3,x_4$ induce a $K_4$ in $G$. The proof is very similar as above. Again we claim:
\begin{equation}\label{K4xinotinDnetfr}
D \cap \{x_1,x_2,x_3,x_4\} = \emptyset.
\end{equation}

\noindent
{\em Proof.}
By symmetry, suppose to the contrary that $x_1 \in D$. Then by the e.d.s.\ property, we have $v_3,v_4,x_2,x_3,x_4 \notin D$. Since $v_3$ and $v_4$ have to be dominated by $D$, let $d_3 \in D$ with $d_3v_3 \in E$ and $d_4 \in D$ with $d_4v_4 \in E$.
By the e.d.s.\ property, $d_3$ and $d_4$ are nonadjacent to the neighbors $v_1,v_2,x_2,x_3,x_4$ of $x_1$. Thus, $d_3 \neq d_4$ since otherwise $x_2,v_3,d_3,v_4,x_4$ would induce a $C_5$ in the chordal graph $G$. This implies $d_3v_4 \notin E$ but now, $v_2,x_2,v_3,d_3,x_3,v_4$ induce a net in $G$ which is a contradiction. Thus, $x_1 \notin D$ and correspondingly, $x_2,x_3,x_4 \notin D$, and claim (\ref{K4xinotinDnetfr}) is shown.
$\diamond$

\medskip

Next we claim:
\begin{equation}\label{K4vinotinDnetfr}
D \cap \{v_1,v_2,v_3,v_4\} = \emptyset.
\end{equation}

\noindent
{\em Proof.}
Without loss of generality, suppose to the contrary that $v_1 \in D$. Then by the e.d.s.\ property, we have $v_2,v_4,x_1,x_2,x_3,x_4 \notin D$. Since $v_2$ and $v_4$ have to be dominated by $D$, let $d_2 \in D$ with $d_2v_2 \in E$ and $d_4 \in D$ with $d_4v_4 \in E$.
Since $d_G(v_2,v_4) > 2$, we have $d_2 \neq d_4$.

Moreover, $d_2x_3 \notin E$ since otherwise, $d_2,v_2,x_1,x_3$ induce a $C_4$ in $G$. This implies $d_2v_3 \notin E$ since otherwise,
$d_2,v_3,x_3,x_1,v_2$ induce a $C_5$ in $G$.

Now, if $d_2x_2 \notin E$ then $d_2,v_2,x_2,v_3,x_1,v_1$ induce a net, and if $d_2x_2 \in E$ then $d_2,x_2,x_1,x_3,v_1,v_4$ induce a net, which is a contradiction.

Thus, $v_1 \notin D$, and correspondingly, $v_2,v_3,v_4 \notin D$, and claim (\ref{K4vinotinDnetfr}) is shown.
$\diamond$

\medskip

Again, let $d_i \in D$ be the $D$-neighbor of $v_i$. By (\ref{K4xinotinDnetfr}) and (\ref{K4vinotinDnetfr}) and the distance properties, we have $d_i \neq v_j,x_j$, $i,j \in \{1,2,3,4\}$.
Next we claim that $d_1,d_2,d_3,d_4$ are pairwise distinct:
\begin{equation}\label{K4d1d2d3d4distinctinDnetfr}
|\{d_1,d_2,d_3,d_4\}|=4.
\end{equation}

\noindent
{\em Proof.} Since $d_G(v_1,v_3) > 2$ and $d_G(v_2,v_4) > 2$, we have $d_1 \neq d_3$ and $d_2 \neq d_4$. Thus, $|\{d_1,d_2,d_3,d_4\}| \ge 2$.

If without loss of generality, $d_1=d_4$, i.e., $d_1v_1 \in E$ and $d_1v_4 \in E$ then, since $d_1,v_1,x_1,x_3,v_4$ do not induce a $C_5$ in $G$, we have $d_1x_1 \in E$ or $d_1x_3 \in E$, and if without loss of generality, $d_1x_1 \in E$ and $d_1x_3 \notin E$ then $d_1,x_1,x_3,v_4$ induce a $C_4$ in $G$. Thus, $d_1x_1 \in E$ and $d_1x_3 \in E$.

This shows that if $d_1v_1 \in E$ and $d_1v_4 \in E$ then $d_2 \neq d_3$, and thus $|\{d_1,d_2,d_3,d_4\}| \ge 3$.

Now assume that $|\{d_1,d_2,d_3,d_4\}| = 3$, i.e., $d_1v_1 \in E$ and $d_1v_4 \in E$, $d_2v_3 \notin E$ and $d_3v_2 \notin E$.
Recall $d_1x_1 \in E$ and $d_1x_3 \in E$. Thus, $d_2x_1 \notin E$, $d_2x_3 \notin E$, $d_3x_1 \notin E$, $d_3x_3 \notin E$.

If $d_2x_2 \notin E$ then $d_2,v_2,x_1,v_1,x_2,v_3$ induce a net in $G$, and if $d_2x_2 \in E$ then $d_3x_2 \notin E$ and thus, $d_3,v_3,x_2,v_2,x_3,v_4$ induce a net in $G$ which is a contradiction. Thus, $d_1,d_2,d_3,d_4$ are pairwise distinct, and claim (\ref{K4d1d2d3d4distinctinDnetfr}) is shown.
$\diamond$

\medskip

If $d_1x_1 \notin E$ and $d_1x_4 \notin E$ then $d_1,v_1,x_1,x_4,v_2,v_4$ induce a net in $G$, and correspondingly by symmetry, a similar statement can be made about  $d_i,x_{i-1},x_i$, $i \neq 1$. Thus, we can assume that for each $i \in \{1,\ldots,4\}$, $d_i$ sees at least one of $x_{i-1},x_i$.

If $d_1x_1 \in E$ and $d_1x_4 \in E$ then clearly, $d_2x_1 \notin E$ and $d_4x_4 \notin E$ and thus, by the above, we can assume that $d_2x_2 \in E$ and
$d_4x_3 \in E$ but now, $d_2,x_2,v_3,x_3,d_3,d_4$ induce a net in $G$.

Thus, assume that $d_1$ is adjacent to exactly one of $x_1,x_4$, say $d_1x_1 \in E$ (which implies $d_2x_1 \notin E$) and $d_1x_4 \notin E$.
By symmetry, this holds for $d_2,d_3,d_4$ as well, i.e., $d_2x_2 \in E$, $d_3x_3 \in E$, and $d_4x_4 \in E$. Then $d_1,x_1,d_2,x_2,d_3,x_3$ induce a net in $G$.

\medskip

Thus, when $x_1,x_2,x_3,x_4$ induce a $K_4$ in $G$, then $G^2$ does not contain a $C_4$ with vertices $v_1,v_2,v_3,v_4$.

\medskip

Now suppose to the contrary that $G^2$ contains $C_k$, $k \ge 5$, say with vertices $v_1,\ldots,v_k$ such that $d_G(v_i,v_{i+1}) \le 2$ and $d_G(v_i,v_j) \ge 3$, $i,j \in \{1,\ldots,k\}$, $|i-j| > 1$ (index arithmetic modulo $k$). By Claim \ref{CkinG2}, we have $d_G(v_i,v_{i+1}) = 2$ for each $i \in \{1,\ldots,k\}$; let $x_i$ be a common neighbor of $v_i,v_{i+1}$. Again, by Claim \ref{CkinG2}, the auxiliary vertices $x_1,\ldots,x_k$ are pairwise distinct and $x_ix_{i+1} \in E$ for each $i \in \{1,\ldots,k\}$.

\medskip

Let $x_i,x_j,x_l$ induce a triangle in $G$. We first claim:

\begin{itemize}
\item[$(i)$] If $j=i+1$ but $|i-l| \ge 2$ and $|j-l| \ge 2$ then $x_i,x_j,x_l,v_i,v_{j+1},v_l$ induce a net in $G$.
\item[$(ii)$] If $|i-j| \ge 2$, $|i-l| \ge 2$, and $|j-l| \ge 2$ then $x_i,x_j,x_l,v_i,v_j,v_l$ induce a net in $G$.
\end{itemize}

Since $G$ is chordal, there is a p.e.o.\ $\sigma$ of $G$, and without loss of generality, assume that $x_1$ is the leftmost vertex of $x_1,\ldots,x_k$ in $\sigma$.
Then $x_2x_k \in E$ since the neighborhood of $x_1$ in $x_2,\ldots,x_k$ is a clique.

\medskip

First assume that $k=5$, and in this case, $x_2x_5 \in E$. Since $x_2,x_3,x_4,x_5$ do not induce a $C_4$ in $G$, we have $x_2x_4 \in E$ or $x_3x_5 \in E$; without loss of generality, assume that $x_2x_4 \in E$. But then, $x_2,x_4,x_5$ induce a triangle as in case $(i)$ of the previous claim, which would lead to a net, which is a contradiction.
Next assume that $k=6$, and in this case, $x_2x_6 \in E$. Then for the cycle $x_2,x_3,x_4,x_5,x_6$ (which is no $C_5$ in $G$), the same argument works as for $k=5$. 
Analogously, for every $k \ge 7$, it can be reduced to the case $k-1$ as for $k=6$.

\medskip

Note that for $k \ge 5$, we do not need the existence of an e.d.s.\ in $G$.

\medskip

Thus, Theorem \ref{EDnetfrchordalG2} is shown.
\qed

\medskip

In a very similar way, we can show:
\begin{theorem}\label{EDextgemfrchordalG2}
If $G$ is an extended-gem-free chordal graph with e.d.s.\ then $G^2$ is chordal.
\end{theorem}

\noindent
{\bf Proof.} Let $G=(V,E)$ be an extended-gem-free chordal graph and assume that $G$ has an e.d.s.\ $D$. We first show that $G^2$ is $C_4$-free:

\medskip

Suppose to the contrary that $G^2$ contains a $C_4$, say with vertices $v_1,v_2,v_3,v_4$ such that $d_G(v_i,v_{i+1}) \le 2$ and $d_G(v_i,v_{i+2}) \ge 3$, $i \in \{1,2,3,4\}$ (index arithmetic modulo 4). By Claim \ref{CkinG2}, we have $d_G(v_i,v_{i+1}) = 2$ for each $i \in \{1,2,3,4\}$; let $x_i$ be a common neighbor of $v_i,v_{i+1}$. By Claim \ref{CkinG2}, $x_i \neq x_j$ for $i \neq j$. Since $G$ is chordal, $x_1,x_2,x_3,x_4$ either induce a diamond or $K_4$ in $G$.

Assume first that $x_1,x_2,x_3,x_4$ induce a diamond in $G$, say with $x_1x_3 \in E$ and $x_2x_4 \notin E$.
We claim:
\begin{equation}\label{diamondxinotinDextgemfr}
D \cap \{x_1,x_2,x_3,x_4\} = \emptyset.
\end{equation}

\noindent
{\em Proof.}
First suppose to the contrary that $x_1 \in D$. Then by the e.d.s.\ property, we have $v_3,v_4,x_2,x_3,x_4 \notin D$. Since $v_3$ and $v_4$ have to be dominated by $D$, let $d_3 \in D$ with $d_3v_3 \in E$ and $d_4 \in D$ with $d_4v_4 \in E$. Clearly, $d_3 \neq x_2,x_3$ and $d_4 \neq x_3,x_4$.
By the e.d.s.\ property, $d_3$ and $d_4$ are nonadjacent to the neighbors $v_1,v_2,x_2,x_3,x_4$ of $x_1$. Thus, $d_3 \neq d_4$ since otherwise $x_1,x_2,v_3,d_3,v_4,x_4$ would induce a $C_6$ in the chordal graph $G$. This implies $d_3v_4 \notin E$ but now, $v_1,x_1,x_3,v_4,x_4,v_2,v_3,d_3$ induce an extended gem which is a contradiction. Thus, $x_1 \notin D$ and correspondingly, $x_3 \notin D$.
Now suppose to the contrary that $x_2 \in D$. Then by the e.d.s.\ property, we have $v_1,v_4,x_1,x_3,x_4 \notin D$. Since $v_1$ and $v_4$ have to be dominated by $D$,
let $d_1 \in D$ with $d_1v_1 \in E$ and $d_4 \in D$ with $d_4v_4 \in E$. Clearly, $d_1 \neq x_1,x_4$ and $d_4 \neq x_3,x_4$. By the e.d.s.\ property, $d_1$ and $d_4$ are nonadjacent to the neighbors $v_2,v_3,x_1,x_3$ of $x_2$. Thus, $d_1v_4 \notin E$ since otherwise $d_1,v_1,x_1,x_3,v_4$ would induce a $C_5$ in the chordal graph $G$, and analogously, $d_4v_1 \notin E$. Now, $d_1,v_1,x_1,v_2,x_2,v_3,x_3,v_4$ induce an extended gem which is a contradiction.
Thus, $x_2 \notin D$ and correspondingly, $x_4 \notin D$, and claim (\ref{diamondxinotinDextgemfr}) is shown.
$\diamond$

\medskip

Next we claim:
\begin{equation}\label{diamondvinotinDextgemfr}
D \cap \{v_1,v_2,v_3,v_4\} = \emptyset.
\end{equation}

\noindent
{\em Proof.}
Without loss of generality, suppose to the contrary that $v_1 \in D$. Then by the e.d.s.\ property, we have $v_2,v_4,x_1,x_2,x_3,x_4 \notin D$. Since $v_2$ and $v_4$ have to be dominated by $D$, let $d_2 \in D$ with $d_2v_2 \in E$ and $d_4 \in D$ with $d_4v_4 \in E$.
Since $d_G(v_2,v_4) > 2$, we have $d_2 \neq d_4$.

Moreover, $d_2x_3 \notin E$ since otherwise, $d_2,v_2,x_1,x_3$ induce a $C_4$ in $G$. This implies $d_2v_3 \notin E$ since otherwise,
$d_2,v_3,x_3,x_1,v_2$ induce a $C_5$ in $G$.
But now, $v_1,x_1,x_3,v_4,x_4,v_3,v_2,d_2$ induce an extended gem which is a contradiction.
Thus, claim (\ref{diamondvinotinDextgemfr}) is shown, i.e., $D \cap \{v_1,v_2,v_3,v_4\} = \emptyset$.
$\diamond$

\medskip

Let $d_i \in D$ be the $D$-neighbor of $v_i$, $i=1,\ldots,4$. By (\ref{diamondxinotinDextgemfr}) and (\ref{diamondvinotinDextgemfr}), we have $d_i \neq v_j,x_j$, $i,j \in \{1,2,3,4\}$.
Next we claim that $d_1,d_2,d_3,d_4$ are pairwise distinct:
\begin{equation}\label{diamondd1d2d3d4distinctinDextgemfr}
|\{d_1,d_2,d_3,d_4\}|=4.
\end{equation}

\noindent
{\em Proof.} Since $d_G(v_1,v_3) > 2$ and $d_G(v_2,v_4) > 2$, we have $d_1 \neq d_3$ and $d_2 \neq d_4$. Thus, $|\{d_1,d_2,d_3,d_4\}| \ge 2$.

If without loss of generality, $d_1=d_4$, i.e., $d_1v_1 \in E$ and $d_1v_4 \in E$ then, since $d_1,v_1,x_1,x_3,v_4$ do not induce a $C_5$ in $G$, we have $d_1x_1 \in E$ or $d_1x_3 \in E$, and if without loss of generality, $d_1x_1 \in E$ and $d_1x_3 \notin E$ then $d_1,x_1,x_3,v_4$ induce a $C_4$ in $G$. Thus, $d_1x_1 \in E$ and $d_1x_3 \in E$.

This shows that if $d_1v_1 \in E$ and $d_1v_4 \in E$ then $d_2 \neq d_3$, and thus $|\{d_1,d_2,d_3,d_4\}| \ge 3$.

Now assume that $|\{d_1,d_2,d_3,d_4\}| = 3$, i.e., $d_1v_1 \in E$ and $d_1v_4 \in E$, $d_2v_3 \notin E$ and $d_3v_2 \notin E$.
Recall $d_1x_1 \in E$ and $d_1x_3 \in E$. Thus, $d_2x_1 \notin E$, $d_2x_3 \notin E$, $d_3x_1 \notin E$, $d_3x_3 \notin E$.
Then $v_1,x_1,x_3,v_4,d_1,v_2,d_2,v_3$ induce an extended gem which is a contradiction.

Thus, $d_1,d_2,d_3,d_4$ are pairwise distinct, and claim (\ref{diamondd1d2d3d4distinctinDextgemfr}) is shown.
$\diamond$

\medskip

If $d_1x_1 \notin E$ and $d_1x_4 \notin E$ then, since $d_1,v_1,x_1,x_2$ do not induce a $C_4$ in $G$, we have $d_1x_2 \notin E$, and accordingly,
since $d_1,v_1,x_4,x_3$ do not induce a $C_4$ in $G$, we have $d_1x_3 \notin E$, but now
$d_1,v_1,x_1,v_2,x_2,v_3,x_3,v_4$ induce an extended gem in $G$ which is a contradiction.

Thus, we can assume that for each $i \in \{1,\ldots,4\}$, $d_i$ sees at least one of $x_{i-1},x_i$ (index arithmetic modulo $4$).

If $d_1x_1 \in E$ and $d_1x_4 \in E$ then clearly, $d_2x_1 \notin E$ and $d_4x_4 \notin E$ and thus, by the above, we can assume that $d_2x_2 \in E$ and
$d_4x_3 \in E$ but now, $v_2,x_1,v_1,x_4,v_4,x_3,v_3,d_3$ induce an extended gem in $G$.

Thus, assume that $d_1$ is adjacent to exactly one of $x_1,x_4$, say $d_1x_1 \in E$ (which implies $d_2x_1 \notin E$) and $d_1x_4 \notin E$.
By symmetry, this holds for $d_2,d_3,d_4$ as well, i.e., $d_2x_2 \in E$, $d_3x_3 \in E$, and $d_4x_4 \in E$. Then
$v_1,x_1,v_2,x_2,v_3,x_3,v_4,d_4$ induce an extended gem in $G$.

\medskip

Now assume that $x_1,x_2,x_3,x_4$ induce a $K_4$ in $G$. The proof is very similar as above. Again we claim:
\begin{equation}\label{K4xinotinDextgemfr}
D \cap \{x_1,x_2,x_3,x_4\} = \emptyset.
\end{equation}

\noindent
{\em Proof.}
By symmetry, suppose to the contrary that $x_1 \in D$. Then by the e.d.s.\ property, we have $v_3,v_4,x_2,x_3,x_4 \notin D$. Since $v_3$ and $v_4$ have to be dominated by $D$, let $d_3 \in D$ with $d_3v_3 \in E$ and $d_4 \in D$ with $d_4v_4 \in E$.
By the e.d.s.\ property, $d_3$ and $d_4$ are nonadjacent to the neighbors $v_1,v_2,x_2,x_3,x_4$ of $x_1$. Thus, $d_3 \neq d_4$ since otherwise $x_2,v_3,d_3,v_4,x_4$ would induce a $C_5$ in the chordal graph $G$. This implies $d_3v_4 \notin E$ but now, $v_2,x_2,x_4,v_1,x_1,v_3,d_3,v_4$ induce an extended gem in $G$ which is a contradiction. Thus, $x_1 \notin D$ and correspondingly, $x_2,x_3,x_4 \notin D$, and claim (\ref{K4xinotinDextgemfr}) is shown.
$\diamond$

\medskip

Next we claim:
\begin{equation}\label{K4vinotinDextgemfr}
D \cap \{v_1,v_2,v_3,v_4\} = \emptyset.
\end{equation}

\noindent
{\em Proof.}
Without loss of generality, suppose to the contrary that $v_1 \in D$. Then by the e.d.s.\ property, we have $v_2,v_4,x_1,x_2,x_3,x_4 \notin D$. Since $v_2$ and $v_4$ have to be dominated by $D$, let $d_2 \in D$ with $d_2v_2 \in E$ and $d_4 \in D$ with $d_4v_4 \in E$.
Since $d_G(v_2,v_4) > 2$, we have $d_2 \neq d_4$.

Moreover, $d_2x_3 \notin E$ since otherwise, $d_2,v_2,x_1,x_3$ induce a $C_4$ in $G$. This implies $d_2v_3 \notin E$ since otherwise,
$d_2,v_3,x_3,x_1,v_2$ induce a $C_5$ in $G$.

Now, $v_1,x_1,x_3,v_4,x_4,v_2,d_2,v_3$ induce an extended gem which is a contradiction.
Thus, $v_1 \notin D$, and correspondingly $v_2,v_3,v_4 \notin D$, and claim (\ref{K4vinotinDextgemfr}) is shown.
$\diamond$

\medskip

Again, let $d_i \in D$ be the $D$-neighbor of $v_i$, $i=1,\ldots,4$. By (\ref{K4xinotinDextgemfr}) and (\ref{K4vinotinDextgemfr}), we have $d_i \neq v_j,x_j$, $i,j \in \{1,2,3,4\}$.
Next we claim that $d_1,d_2,d_3,d_4$ are pairwise distinct:
\begin{equation}\label{K4d1d2d3d4distinctinDextgemfr}
|\{d_1,d_2,d_3,d_4\}|=4.
\end{equation}

\noindent
{\em Proof.} Since $d_G(v_1,v_3) > 2$ and $d_G(v_2,v_4) > 2$, we have $d_1 \neq d_3$ and $d_2 \neq d_4$. Thus, $|\{d_1,d_2,d_3,d_4\}| \ge 2$.

If without loss of generality, $d_1v_1 \in E$ and $d_1v_4 \in E$ then, since $d_1,v_1,x_1,x_3,v_4$ do not induce a $C_5$ in $G$, we have $d_1x_1 \in E$ or
$d_1x_3 \in E$, and if $d_1x_1 \in E$ and $d_1x_3 \notin E$ then $d_1,x_1,x_3,v_4$ induce a $C_4$ in $G$. Thus, $d_1x_1 \in E$ and $d_1x_3 \in E$.

This shows that if $d_1v_1 \in E$ and $d_1v_4 \in E$ then $d_2 \neq d_3$, and thus $|\{d_1,d_2,d_3,d_4\}| \ge 3$.

Now assume that $|\{d_1,d_2,d_3,d_4\}| = 3$, i.e., $d_1v_1 \in E$ and $d_1v_4 \in E$, $d_2v_3 \notin E$ and $d_3v_2 \notin E$. Since $d_1x_1 \in E$ and $d_1x_3 \in E$,
 $v_1,x_1,x_3,v_4,d_1,v_2,d_2,v_3$ induce an extended gem in $G$ which is a contradiction.
Thus, $d_1,d_2,d_3,d_4$ are pairwise distinct, and claim (\ref{K4d1d2d3d4distinctinDextgemfr}) is shown.
$\diamond$

\medskip

If $d_1x_1 \notin E$ then, since $d_1,v_1,x_1,x_2$ do not induce a $C_4$ in $G$, we have $d_1x_2 \notin E$, and analogously, $d_1x_3 \notin E$. But now
$v_2,x_1,x_3,v_3,x_2,v_1,d_1,v_4$ induce an extended gem in $G$ which is a contradiction.
Thus, $d_1x_1 \in E$ and by symmetry, $d_1x_4 \in E$ but now, by the e.d.s.\ property, $d_2x_1 \notin E$ and $d_2x_4 \notin E$, and since $d_2,v_2,x_1,x_3$ do not induce a $C_4$, we have $d_2x_3 \notin E$. But now, $v_1,x_1,x_3,v_4,x_4,v_2,d_2,v_3$ induce an extended gem in $G$ which is a contradiction.

\medskip

Thus, when $x_1,x_2,x_3,x_4$ induce a diamond or $K_4$ in $G$, then $G^2$ does not contain a $C_4$ with vertices $v_1,v_2,v_3,v_4$.

\medskip

Now suppose to the contrary that $G^2$ contains $C_k$, $k \ge 5$, say with vertices $v_1,\ldots,v_k$ such that $d_G(v_i,v_{i+1}) \le 2$ and $d_G(v_i,v_j) \ge 3$, $i,j \in \{1,\ldots,k\}$, $|i-j| > 1$ (index arithmetic modulo $k$). By Claim \ref{CkinG2}, we have $d_G(v_i,v_{i+1}) = 2$ for each $i \in \{1,\ldots,k\}$; let $x_i$ be a common neighbor of $v_i,v_{i+1}$. Again, by Claim \ref{CkinG2}, the auxiliary vertices $x_1,\ldots,x_k$ are pairwise distinct and $x_ix_{i+1} \in E$ for each $i \in \{1,\ldots,k\}$.

\medskip

Clearly, since $G$ is chordal, there is an edge $x_ix_{i+2} \in E$. We claim:
\begin{equation}\label{xixi+2inExi+1notinDextgemfr}
\mbox{If } x_ix_{i+2} \in E \mbox{ then } x_i,x_{i+1},x_{i+2} \notin D \mbox{ and } v_{i+1},v_{i+2} \notin D.
\end{equation}

\noindent
{\em Proof.} Without loss of generality, let $x_1x_3 \in E$. If $x_2 \in D$ then clearly, $v_1 \notin D$ and $x_k,x_1 \notin D$; let $d_1 \in D$ be a new vertex with $d_1v_1 \in E$. Clearly, $d_1 \cojoin \{x_1,v_2,x_2,v_3,x_3,v_4\}$ but now, $x_1,v_2,x_2,v_3,x_3,v_4,v_1,d_1$ induce an extended gem. Thus, $x_2 \notin D$.

If $x_1 \in D$ then clearly, $v_4 \notin D$ and $x_3,x_4 \notin D$; let $d_4 \in D$ be a new vertex with $d_4v_4 \in E$. Clearly, $d_4 \cojoin \{v_1,x_1,v_2,x_2,v_3,x_3\}$ but now,
$v_1,x_1,v_2,x_2,v_3,x_3,v_4,d_4$ induce an extended gem. Thus, $x_1 \notin D$ and correspondingly, $x_3 \notin D$ by symmetry.

If $v_2 \in D$ then clearly, $v_1 \notin D$ and $x_k,x_1 \notin D$; let $d_1 \in D$ be a new vertex with $d_1v_1 \in E$. As before, $d_1 \cojoin \{x_1,v_2,x_2,v_3,x_3,v_4\}$ but now,
$d_1,v_1,x_1,v_2,x_2,v_3,x_3,v_4$ induce an extended gem. Thus, $v_2 \notin D$ and correspondingly, $v_3 \notin D$ by symmetry
which shows (\ref{xixi+2inExi+1notinDextgemfr}).
$\diamond$

\medskip

Next we claim:
\begin{equation}\label{xixi+2notxi+2xi+4extgemfr}
\mbox{If } x_ix_{i+2} \in E \mbox{ then } x_{i+2}x_{i+4} \notin E \mbox{ and } x_{i-2}x_{i} \notin E.
\end{equation}

\noindent
{\em Proof.} Without loss of generality, let $x_1x_3 \in E$ and suppose to the contrary that $x_3x_5 \in E$. Then by (\ref{xixi+2inExi+1notinDextgemfr}),
there are new vertices $d_3,d_4,d_5 \in D$, $d_3,d_4,d_5 \notin \{v_3,v_4,v_5,x_2,x_3,x_4,x_5\}$, with $d_3v_3 \in E$, $d_4v_4 \in E$ and $d_5v_5 \in E$.
We first claim that $d_3 \neq d_4$:

\medskip

Suppose to the contrary that $d_3=d_4$.
If $x_2x_4 \in E$ then, since $d_3,v_3,x_2,x_4,v_4$ do not induce a chordless cycle, we have $d_3x_2 \in E$ and $d_3x_4 \in E$, but now, $v_3,x_2,x_4,v_4,d_3,v_2,v_5,d_5$ induce an extended gem. Thus, let $x_2x_4 \notin E$.

Since $v_2,x_1,x_3,v_3,x_2,v_1,x_4,v_5$ do not induce an extended gem, we have $x_1x_4 \in E$.
Since $d_3,v_3,x_2,x_1,x_4,v_4$ do not induce a chordless cycle, we have $d_3x_2 \in E$, $d_3x_1 \in E$, and $d_3x_4 \in E$. Thus, by the e.d.s.\ property,
$d_5x_1 \notin E$, $d_5x_4 \notin E$, and thus, $d_5v_1 \notin E$ since $d_5,v_1,x_1,x_4,v_5$ do not induce a $C_5$. But now,
$x_2,x_1,x_4,v_4,d_3,v_1,v_5,d_5$ induce an extended gem which is a contradiction. Thus, $d_3 \neq d_4$ is shown.

\medskip

By the e.d.s.\ property, $d_3x_3 \notin E$ or $d_4x_3 \notin E$. Recall that $x_3x_5 \in E$ was supposed, and thus, say without loss of generality, $d_4x_3 \notin E$. Then by the chordality of $G$, $d_4x_2 \notin E$ and $d_4x_1 \notin E$, and clearly, $d_4 \cojoin \{v_1,v_2,v_3\}$ but now,
$v_1,x_1,v_2,x_2,v_3,x_3,v_4,d_4$ induce an extended gem. Thus, (\ref{xixi+2notxi+2xi+4extgemfr}) is shown.
$\diamond$

\medskip

For a $C_5$ in $G^2$, fact (\ref{xixi+2notxi+2xi+4extgemfr}) leads to a $C_4$ in $G$ induced by $x_1,x_3,x_4,x_5$ if $x_1x_3 \in E$.
Thus, from now on, let $k \ge 6$. We claim:

\begin{equation}\label{xixi+2notxi+1xi+3extgemfr}
\mbox{If } x_ix_{i+2} \in E \mbox{ then } x_{i+1}x_{i+3} \notin E \mbox{ and } x_{i-1}x_{i+1} \notin E.
\end{equation}

\noindent
{\em Proof.}
Without loss of generality, let $x_1x_3 \in E$ and suppose to the contrary that $x_2x_4 \in E$. Then by (\ref{xixi+2notxi+2xi+4extgemfr}),
$x_3x_5 \notin E$ and $x_4x_6 \notin E$ as well as $x_1x_{k-1} \notin E$ and $x_2x_k \notin E$,
and since $G$ is chordal, $x_3x_6 \notin E$ and $x_2x_{k-1} \notin E$.

Since $v_2,x_2,v_3,x_3,v_4,x_4,x_5,v_6$ does not induce an extended gem, we have $x_2x_5 \in E$. For $k=6$ this contradicts the fact that $x_2x_{k-1} \notin E$, i.e., $x_2x_5 \notin E$. Thus, from now on, let $k \ge 7$.

Since $v_2,x_2,x_3,x_4,v_5,x_5,x_6,v_7$ do not induce an extended gem, we have $x_2x_6 \in E$ (recall $x_3x_5 \notin E$, $x_3x_6 \notin E$ and $x_4x_6 \notin E$). For $k=7$, this implies that
$x_1,x_2,x_6,x_7$ induce a $C_4$ which is a contradiction. Thus, let $k \ge 8$ but now, $x_2,v_3,x_3,v_4,x_4,v_5,x_6,v_6$ induce an extended gem. Thus, (\ref{xixi+2notxi+1xi+3extgemfr}) is shown.
$\diamond$

\medskip

Recall that $k \ge 6$; without loss of generality, let $x_1x_3 \in E$. Then by (\ref{xixi+2notxi+2xi+4extgemfr}) and (\ref{xixi+2notxi+1xi+3extgemfr}), we have
$x_2x_4 \notin E$, $x_kx_2 \notin E$, and $x_3x_5 \notin E$, $x_{k-1}x_1 \notin E$.
Since $G$ is chordal, we have $x_2x_5 \notin E$.

Since $v_2,x_1,x_3,v_3,x_2,x_4,v_5,v_1$ do not induce an extended gem, we have $x_1x_4 \in E$.

Since $x_2,x_1,x_4,v_4,x_3,x_5,v_6,v_1$ do not induce an extended gem, we have $x_1x_5 \in E$ (which, for $k=6$ contradicts the fact that $x_{k-1}x_1 \notin E$) but now, $v_2,x_1,x_3,v_3,x_2,x_5,v_5,v_4$ induce an extended gem.

\medskip

Thus, Theorem \ref{EDextgemfrchordalG2} is shown.
\qed

\medskip

In the case of net-free chordal graphs, Theorem \ref{EDnetfrchordalG2} generalizes the corresponding result for AT-free chordal graphs (i.e., interval graphs---see e.g.\ \cite{BraLeSpi1999}).

\medskip

By \cite{BraFicLeiMil2013}, and since MWIS is solvable in linear time for chordal graphs \cite{Frank1975}, we obtain:
\begin{corollary}\label{WEDnetfrchordalpol}
WED is solvable in time ${\cal O}(n^3)$ for net-free chordal graphs and for extended-gem-free chordal graphs.
\end{corollary}

Theorems \ref{EDnetfrchordalG2} and \ref{EDextgemfrchordalG2} and the subsequent lemma imply further polynomial cases for WED:

\begin{lemma}[\cite{BraGia2014,BraGiaMil2018}]\label{P2cojoin}
If WED is solvable in polynomial time for $F$-free graphs then WED is solvable in polynomial time for $(P_2+F)$-free graphs.
\end{lemma}

This clearly implies the corresponding fact for $(P_1+F)$-free graphs.

\medskip

Recall Lemma \ref{butterflyNPc} for $H \in $ $\{2P_3$, $K_3+P_3$, $2K_3$, butterfly, extended butterfly, extended co-$P$, extended chair, double-gem$\}$. Now we consider induced subgraphs $H'=H-x$ of $H$ which are the following:

\begin{itemize}
\item[$-$] $H=2P_3$: $H' \in \{P_2+P_3,P_3+2P_1\}$
\item[$-$]  $H=K_3+P_3$: $H' \in \{P_2+P_3,K_3+P_2,K_3+2P_1\}$
\item[$-$]  $H=2K_3$: $H' = P_2+K_3$
\item[$-$]  $H=$ butterfly: $H' \in \{2P_2$,paw$\}$
\item[$-$]  $H=$ extended butterfly: $H' \in \{K_3+P_2$,co-P$\}$
\item[$-$]  $H=$ extended co-P: $H' \in \{K_3+P_2,P_5$,paw$+P_1$,co-P$\}$
\item[$-$]  $H=$ extended chair: $H' \in \{K_3+2P_1,P_2+P_3$,chair,co-P$\}$
\item[$-$]  $H=$ double-gem: $H' \in \{$co-P,gem$\}$
\end{itemize}

\begin{corollary}\label{coPK3P2frcorollary}
For every proper induced subgraph $H'$ of any graph $H \in $ $\{2P_3$, $K_3+P_3$, $2K_3$, butterfly, extended butterfly, extended co-$P$, extended chair, double-gem$\}$, WED is solvable in polynomial time for $H'$-free chordal graphs.
\end{corollary}

\noindent
{\bf Proof.}
By \cite{BraDabHuaPau2015}, the clique-width of co-chair-free chordal graphs is bounded, and by \cite{GolRot2000}, the clique-width of gem-free chordal graphs is bounded. By Theorem \ref{EDextgemfrchordalG2}, WED is solvable in polynomial time for chair-free chordal graphs since chair is an induced subgraph of extended gem, and similarly, for co-$P$-free chordal graphs. By Lemma \ref{P2cojoin}, WED is solvable in polynomial time for $(K_3+P_2)$-free chordal graphs and since the clique-width of $K_3$-free chordal graphs is bounded.
In all other cases, we can use Lemma \ref{P2cojoin} and the fact that WED is solvable in polynomial time (even in linear time) for $P_5$-free graphs (and thus also for $2P_2$-free graphs).
\qed

\section{WED for $S_{1,2,3}$-Free Chordal Graphs - a Direct Approach}\label{WEDS123frchordalgr}

By Lemma \ref{butterflyNPc}, and since $S_{1,1,4}$ as well as $S_{1,3,3}$ contain $2P_3$ as an induced subgraph,
WED is \NP-complete for $S_{1,1,4}$-free chordal as well as for $S_{1,3,3}$-free chordal graphs. In this section, we give a polynomial-time solution for WED on $S_{1,2,3}$-free chordal graphs by a direct approach.

This generalizes WED for $S_{1,2,2}$-free chordal graphs as well as for $S_{1,1,3}$-free chordal graphs ($S_{1,2,2}$ and $S_{1,1,3}$ are induced subgraphs of extended gem---see Figure \ref{netextgem} and recall Theorem \ref{EDextgemfrchordalG2})
and for $P_6$-free chordal graphs (recall \cite{BraEscFri2015,BraEscFriKar2017}).

\medskip

Throughout this section, let $G=(V,E)$ be a prime $S_{1,2,3}$-free chordal graph; recall that WED for $G$ can be reduced to prime graphs
\cite{BraGia2014,BraGiaMil2018,BraMilNev2013}. For any vertex $v \in V$, let
\begin{itemize}
\item[ ] $Z^+(v) := \{u \in V: N[v] \subset N[u]\}$, and
\item[ ] $Z^-(v) := \{u \in V: N[u] \subset N[v]\}$.
\end{itemize}

Let us say that a vertex $v \in V$ is a {\em maximal vertex} of $G$ if $Z^+(v)=\emptyset$. Clearly, $G$ has at least one maximal vertex.

\begin{lemma}\label{WEDmaxvdirectS123frchordalgr}
Let $v \in V$ be a maximal vertex of $G$. Then a minimum $($finite$)$ weight e.d.s.\ $D$ with $v \in D$ $($if $D$ exists$)$ can be computed in polynomial time.
\end{lemma}

\noindent
{\bf Proof.} Assume that $D$ is a (possible) e.d.s.\ of finite weight of $G$ with $v \in D$. Recall that $G$ is prime (and thus, connected); then, by excluding the trivial case in which $V = \{v\}$, $G$ is not a clique. As usual, let $N_0=\{v\}$ and let $N_1,\ldots,N_t$ (for some natural $t$) denote the distance levels of $v$ in $G$. Then $N_0, N_1,\ldots,N_t\}$ is a partition of $V$.
Clearly, since $v \in D$, $(N_1 \cup N_2) \cap D = \emptyset$.
Since $G$ is chordal, we have:

\begin{Claim}\label{xNicliqueN(x)Ni-1}
For every $i \in \{1,\ldots,t\}$ and every vertex $x \in N_i$, $N(x) \cap N_{i-1}$ is a clique, and in particular, $x$ contacts exactly one component of $G[N_{i-1}]$.
\end{Claim}

\begin{Claim}\label{u1inN1nonadjz1}
\mbox{ }
\begin{itemize}
\item[$(i)$] For any vertex $u_1 \in N_1$, there is a vertex $z_1 \in N_1$ with $z_1u_1 \notin E$.
\item[$(ii)$] For any vertex $u_2 \in N_2$, with neighbor $u_1 \in N_1$, there is a vertex $z_1 \in N_1$ with $z_1u_1 \notin E$ and $z_1u_2 \notin E$.
\item[$(iii)$] For any vertex $u_i \in N_i$, $i \ge 2$, there is a chordless path $P_{u_iv}$ with at least four vertices including $u_i$ and $v$.
\end{itemize}
\end{Claim}

\noindent
{\em Proof.}
Statement $(i)$ holds since $v$ is a maximal vertex of $G$ and since the prime graph $G$ is not a clique. Statement $(ii)$ holds by $(i)$ and since $G$ is chordal.
If $i \ge 3$ then statement $(iii)$ trivially holds by construction. If $i=2$ then it easily follows by $(i)$ and $(ii)$.
$\diamond$

\begin{Claim}\label{P3}
For any fixed $i$, $i \in \{2,\ldots,t-1\}$, let
\begin{itemize}
\item[ ] $X := \{x \in N_i: x$ has a neighbor in $D \cap N_{i+1}\}$, let
\item[ ] ${\cal C}_X := \{Y_1,\ldots,Y_q\}$ $($for some natural $q$ $)$ be the family of connected components of $G[N_{i+1}]$ contacting~$X$, and let
\item[ ] $X_i:=\{x \in X: x$ contacts $Y_i\}$, $i=1,\ldots,q$.
\end{itemize}

Then the following statements hold:

\begin{enumerate}
\item [$(i)$] For every $x \in X$, $x$ contacts exactly one of $Y_1,\ldots,Y_q$, and thus, for $i \neq j$, $X_i \cap X_j = \emptyset$, i.e.,
$X$ admits a partition $\{X_1,\ldots,X_q\}$ such that for $h,k \in \{1,\ldots,q\}$, $k \neq h$, $Y_h$ contacts $X_h$ and does not contact $X_k$.
\item[$(ii)$] For every $h \in \{1,\ldots,q\}$, $|D \cap Y_h| = 1$, say $D \cap Y_h = \{d_h\}$, and $d_h$ dominates $X_h \cup Y_h$, i.e.,
$X_h \cup Y_h \subseteq N[d_h]$.
\end{enumerate}
\end{Claim}

\noindent
{\em Proof.}
$(i)$: First we prove that for any $x \in X$, $x$ contacts exactly one of $Y_1,\ldots,Y_q$: Without loss of generality, suppose to the contrary that $x$ contacts
$Y_1$ and $Y_2$, and assume that the neighbor of $x$ in $D \cap N_{i+1}$, say $d$, belongs to $Y_1$.
Then let $y$ be a neighbor of $x$ in $Y_2$: By the e.d.s.\ property, $y$ has a neighbor in $D$, say $d'$, with $d' \neq d$. Clearly, by the e.d.s.\ property and by definition of $X$, we have $d' \notin X$ and $xd' \notin E$ and thus, by Claim \ref{xNicliqueN(x)Ni-1}, $d' \notin N_i$.

Thus, $d' \in N_{i+1} \cup N_{i+2}$. Then $d',y,d,x$, and three further vertices of the path $P_{xv}$ found by Claim \ref{u1inN1nonadjz1} $(iii)$ induce an $S_{1,2,3}$, which is a contradiction.

\medskip

Thus, for $i \neq j$, $X_i \cap X_j = \emptyset$, and $(i)$ follows directly by the above and by definition of $X,X_i$ and ${\cal C}_X$.

\medskip

\noindent
$(ii)$: First we prove that $|D \cap Y_h| = 1$ (note that $D \cap Y_h \neq \emptyset$, by the proof of statement ($i$) of this claim: Suppose to the contrary that there are $d,d' \in D \cap Y_h$, $d \neq d'$. Since $G$ is connected and by definition of $X$, there are $x \in X$ with $xd \in E$ and $x' \in X$ with $x'd' \in E$. By the e.d.s.\ property, the shortest path, say $P$, in $Y_h$ from $d$ to $d'$ has at least two internal vertices, i.e., there exist $a,b \in P$ with $da \in E$ and $bd' \in E$. Since $G$ is $S_{1,2,3}$-free, by Claim \ref{u1inN1nonadjz1} $(iii)$ and by the e.d.s.\ property, $x$ is nonadjacent to all vertices of $P \setminus \{d,a\}$, while $x'$ is nonadjacent to all vertices of $P \setminus \{b,d'\}$, which contradicts the fact that $G$ is chordal. Thus, $|D \cap Y_h| = 1$;
let $D \cap Y_h = \{d_h\}$.

\medskip

Next we claim that $d_h$ dominates $X_h$: This follows by definition of $X$, by statement $(i)$ of this claim, and by the e.d.s.\ property. By the way, by Claim \ref{xNicliqueN(x)Ni-1}, $X_h$ is a clique.

\medskip

Finally we claim that $d_h$ dominates $Y_h$: Suppose to the contrary that there is a vertex $y \in Y_h$ with $yd_h \notin E$.
Since $D \cap Y_h = \{d_h\}$, we have $y \notin D$. Then there is $d \in D$, $d \neq d_h$, with $yd \in E$. Let $P'$ be a shortest path in $Y_h$ between $d_h$ and $y$, and let $x \in X$ be adjacent to $d_h$ (by the above, $d_h$ dominates $X_h$). Clearly, by the e.d.s.\ property, $xd \notin E$.

\medskip

If $xy \in E$ then by Claim \ref{xNicliqueN(x)Ni-1}, $d \not \in N_{i}$, i.e., $d \in N_{i+1} \cup N_{i+2}$; then $d,y,d_h,x$, and three further vertices of the path $P_{xv}$ found by Claim \ref{u1inN1nonadjz1} $(iii)$ induce an $S_{1,2,3}$ which is a contradiction. Thus $xy \notin E$.

\medskip

If $d \in N_{i}$ then, by considering the (not necessarily induced) path formed by vertices $x,d_h,P',y,d$, we get a contradiction to the fact that $G$ is chordal. Thus, $d \in N_{i+1} \cup N_{i+2}$. Then let $y'$ be a neighbor of $y$ in $N_i$; clearly, by the e.d.s.\ property, $y' \not \in D$.

Note that $y'd_h \notin E$ (else $d_h,y',y,d$, and three further vertices of the path $P_{y'v}$ found by Claim \ref{u1inN1nonadjz1} $(iii)$ induce an $S_{1,2,3}$)
and $y'x \in E$, else by considering the (not necessarily induced) path formed by $x,d_h$,$P',y,y'$, we get a contradiction since $G$ is chordal.

Then there is $d' \in D$ adjacent to $y'$. Clearly, $d' \neq d_h$ by the above. Furthermore $d' \neq d$: Otherwise, if $y'd \in E$ then $d \in N_{i+1}$, and then by considering the path between $d_h$ and $d$ in $N_{i+1}$ (consisting of path $P'$ in $Y_{h}$ between $d_h$ and $y$ and additionally $d$) we get a contradiction to the fact that $G$ is chordal by an argument similar to the one above for showing that $|D \cap Y_h| = 1$.

\medskip

If $d' \in N_{i-1}$ then, since $D \cap (N_1 \cup N_2) = \emptyset$, $i \ge 4$, and $d_h,x,y,y'$, and three further vertices of the path $P_{y'v}$ found by Claim \ref{u1inN1nonadjz1} $(iii)$ containing $d'$ induce an $S_{1,2,3}$ which is a contradiction.

\medskip

If $d' \in N_{i}$ then $i \ge 3$ since $D \cap (N_1 \cup N_2) = \emptyset$. Since $G$ is chordal, $y'$ and $d'$ have a common neighbor in $N_{i-1}$, say $z$, and then $zx \in E$ since otherwise $d_h,x,y,y'$, and three further vertices of the path $P_{y'v}$ found by Claim \ref{u1inN1nonadjz1} $(iii)$ containing $z$ induce an $S_{1,2,3}$.
Now, since $xz \in E$, the vertices $d',d_h,x,z$, and three further vertices of the path $P_{zv}$ found by Claim \ref{u1inN1nonadjz1} $(iii)$ induce an $S_{1,2,3}$, which is a contradiction.

\medskip

Finally if $d' \in N_{i+1}$ then $d,y,d',y'$, and three further vertices of the path $P_{y'v}$ found by Claim \ref{u1inN1nonadjz1} $(iii)$ induce an $S_{1,2,3}$, which is a contradiction.

\medskip

Thus, Claim \ref{P3} is shown.
$\diamond$

\begin{Claim}\label{P4}
For every component $K$ of $G[N_i]$, $i \in \{3,\ldots,t\}$, we have
\begin{enumerate}
\item [$(i)$] $|D \cap K| \leq 1$, and
\item[$(ii)$] if $|D \cap K| = 1$, say $D \cap K=\{d\}$ then $d$ dominates $K$.
\end{enumerate}
\end{Claim}

\noindent
{\em Proof.}
$(i)$: It can be proved similarly to the first paragraph of the proof of Claim \ref{P3} $(ii)$.

\medskip

\noindent
$(ii)$: It follows by Claim \ref{P3} $(ii)$ since $d$ (and thus $K$) contacts a set of vertices of $N_{i-1}$ which consequently have a neighbor in $D \cap N_{i}$.
$\diamond$

\medskip

Now let us consider the problem of checking whether such an e.d.s.\ $D$ of $G$  with $v \in D$ does exist.
According to Claim \ref{xNicliqueN(x)Ni-1}, graph $G$ can be viewed as a tree $T$ rooted at $\{v\}$, whose nodes are the connected components of $G[N_i]$ for $i \in  \{0,1,\ldots,t\}$ (recall $N_0 := \{v\}$), such that two nodes are adjacent if and only if the corresponding connected components contact each other.

Then for any connected component $K$ of $G[N_i]$, $i \in \{0,1,\ldots,t\}$, let $T(K)$ denote the vertex set of the induced subgraph of $G$ corresponding to the subtree of $T$ rooted at $K$. In particular $N_0$ has a unique connected component (recall $N_0 := \{v\}$), say $K_0$, so that $T(K_0) = V$.

According to Claim \ref{P4}, let us say that a vertex $d$ of $G$ of finite weight, belonging to a connected component say $K$ of $G[N_i]$, $i \in \{0,1,\ldots,t\}$, is a {\em $D$-candidate} (or equivalently let us say that $K$ admits a {\em $D$-candidate} $d$) if
\begin{enumerate}
\item [$(i)$] $d$ dominates $K$, and
\item[$(ii)$] there is an e.d.s.\ in $G[T(K)]$ containing $d$.
\end{enumerate}

\begin{Claim}\label{P5}
An e.d.s.\ $D$ of $G$ with $v \in D$ does exist if and only if $v$ is a $D$-candidate.
\end{Claim}

\noindent
{\em Proof.} It directly follows by the above.
$\diamond$

\begin{Claim}\label{P6}
Let $K$ be a connected component of $G[N_i]$, for any fixed $i \in \{1,\ldots,t\}$, and let $d \in V(K)$ be a vertex of finite weight. Then let $H_j := T(K) \cap N_j$ for $i+1 \le j \le t$, and let

\begin{itemize}
\item[] $A := \{x \in H_{i+1}: xd \notin E\}$;
\item[] ${\cal C}_A=\{A'_1,\ldots,A'_q\}$ be the family of connected components of $G[H_{i+2}]$ contacting $A$;
\item[] $B$ be the vertex set of connected components of $G[H_{i+2}]$ not contacting $A$;
\item[] ${\cal C}_B=\{B'_1,\ldots,B'_{q'}\}$ be the family of connected components of $G[H_{i+3}]$ contacting $B$.
\end{itemize}

Then the following statements hold:

\begin{itemize}
\item[$(i)$] If $A = B = \emptyset$ then $d$ is a $D$-candidate if and only if $d$ dominates $K$.

\item[$(ii)$] If $A \neq \emptyset$ and $B = \emptyset$ then $d$ is a $D$-candidate if and only if $d$ dominates $K$, Claim $\ref{P3}$ $(i)$ holds for $A$ and for ${\cal C}_A$, and according to the notation of Claim $\ref{P3}$, $A$ admits a partition $\{A_1,\ldots,A_q\}$, and each member $A'_h$ of ${\cal C}_A$ admits a $D$-candidate which dominates $A_h \cup A'_h$ and does not contact $N(d) \cap H_{i+1}$.

\item[$(iii)$] If $A = \emptyset$ and $B \neq \emptyset$ then $B$ admits a partition $\{B_1,\ldots,B_q\}$, $d$ is a $D$-candidate if and only if $d$ dominates $K$, Claim $\ref{P3}$ $(i)$ holds for $B$ and for ${\cal C}_B$, and according to the notation of Claim $\ref{P3}$,  
each member $B'_h$ of ${\cal C}_B$ admits a $D$-candidate which dominates $B_h \cup B'_h$.

\item[$(iv)$] If $A \neq \emptyset$ and $B \neq \emptyset$ then $d$ is a $D$-candidate if and only if $d$ dominates $K$, Claim $\ref{P3}$ $(i)$ holds for $A$ and for ${\cal C}_A$, and according to the notation of Claim $\ref{P3}$, each member $A'_h$ of ${\cal C}_A$ admits a $D$-candidate which dominates $A_h \cup A'_h$ and does not contact $N(d) \cap H_{i+1}$, Claim $\ref{P3}$ $(i)$ holds for $B$ and for ${\cal C}_B$, and according to the notation of Claim $\ref{P3}$, each member $B'_h$ of ${\cal C}_B$ admits a $D$-candidate which dominates $B_h \cup B'_h$.
\end{itemize}
\end{Claim}

\noindent
{\em Proof.} It follows by definition of $D$-candidate, by the e.d.s.\ property, by Claim \ref{P3}, and by Claim \ref{P4}; in particular by construction, each vertex of $A$ contacts $V(K) \setminus \{d\}$, each vertex of $B$ contacts $N(d) \cap H_{i+1}$ and no member of ${\cal C}_A$, and then each member of ${\cal C}_A$ contacts no member of ${\cal C}_B$ by Claim \ref{xNicliqueN(x)Ni-1}.
$\diamond$

\medskip

Then by Claims \ref{P5} and \ref{P6}, one can check if e.d.s.\ $D$ with $v \in D$ does exist by the following procedure which can be executed in polynomial time:

\begin{proc}[$v$-Maximal-WED]\label{vMaxWED}

\begin{tabbing}	
xxxxxxx \= \kill\\
{\bf Input:} \> A maximal vertex $v$ of $G$.\\
{\bf Task:} \> A minimum weight e.d.s.\ $D$ of $G$ containing $v$ $($if it exists$)$.
\end{tabbing}

\noindent
{\bf begin}
\begin{itemize}
\item[ ] Let $N_0,N_1,\ldots,N_t$ $($for some natural $t)$, with $N_0 = \{v\}$, be the distance levels of $v$ in $G$.
\item[ ] {\bf for} $i = t, t-1, \ldots, 1,0$ {\bf do}
\item[ ] {\bf begin}
\begin{itemize}
\item[ ] for each component $K$ of $G[N_i]$, detect all $D$-candidates in $K$, and for each $D$-candidate in $K$, say $u$, store (iteratively by the possible $D$-candidates in $C_A$ and in $C_B$) any minimum weight e.d.s.\ of $G[T(K)]$ containing $u$;
\end{itemize}
\item[ ] {\bf end}
\item[ ] {\bf if} $v$ is a $D$-candidate {\bf then} return ``$D$ does exist''
\item[ ] {\bf else} return ``$D$ does not exist''.
\end{itemize}
{\bf end}
\end{proc}



This completes the proof of Lemma \ref{WEDmaxvdirectS123frchordalgr}.
\qed

\begin{theorem}\label{WEDdirectS123frchordalgr}
For $S_{1,2,3}$-free chordal graphs, WED is solvable in polynomial time.
\end{theorem}

\noindent
{\bf Proof.}
Let us observe that, if all vertices of $G$ are maximal, then by Lemma~\ref{WEDmaxvdirectS123frchordalgr}, the WED problem can be solved for $G$ by computing a minimum finite weight e.d.s.\ $D$ with $v \in D$ (if $D$ exists), for all $v \in V$.

\medskip

Then let us focus on those vertices $x$ which are not maximal, i.e., there is a vertex $y$ with $N[x] \subset N[y]$ (which means $x \in Z^-(y)$). Thus, there is a maximal vertex $v$ such that $x \in Z^-(v)$. In particular removing such maximal vertices $v$ leads to new maximal vertices in the reduced graph. Recall that for any graph $G=(V,E)$ and any e.d.s.\ $D$ of $G$, $|D \cap N[x]| = 1$ for every $x \in V$.

\medskip

\noindent
{\bf Fact 1.} {\em Let $v \in V$ be a maximal vertex of $G$, with $Z^-(v) \neq \emptyset$, and let $x \in Z^-(v)$. If $G$ has an e.d.s., say $D$, then $D \cap (N(v) \setminus N(x))=\emptyset$.

Define a reduced weighted graph $G^*$ from $G$ as follows:

\begin{itemize}
\item[$(i)$] For each vertex $x \in Z^-(v)$, assign weight $\infty$ to all vertices in $N(v) \setminus N(x)$, and
\item[$(ii)$] remove $v$, i.e., $V(G^*)=V \setminus \{v\}$ (and reduce $G^*$ to its prime connected components; recall that WED can be reduced to prime graphs).
\end{itemize}
Then the problem of checking if $G$ has a finite (minimum weight) e.d.s.\ not containing $v$ can be reduced to that of checking if $G^*$ has a finite (minimum weight) e.d.s.}

\medskip

\noindent
{\em Proof.}
The reduction is correct by the e.d.s.\ property and by definition of $Z^-(v)$. Moreover, by the e.d.s. property, by definition of $Z^-(v)$ and by construction of $G^*$, every (possible) e.d.s.\ of finite weight of $G^*$ contains exactly one vertex which is a neighbor of $v$ in $G$ since $|D \cap N[x]| = 1$ for a vertex $x \in Z^-(v)$.
$\diamond$

\medskip

Since the above holds in a hereditary way for any subgraph of $G$, and since WED for any graph $H$ can be reduced to the same problem for the connected components of $H$, let us introduce a possible algorithm to solve WED for $G$ in polynomial time.

\begin{algo}[WED-$S_{1,2,3}$-Free-Chordal-Graphs]\label{algWEDS123frchgr}

\begin{tabbing}	
xxxxxxx \= \kill\\
{\bf Input:} \> Graph $G = (V,E)$.\\
{\bf Task:} \>  A minimum (finite) weight e.d.s.\ of $G$ (if it exists).
\end{tabbing}

\noindent
{\bf begin}
\begin{itemize}
\item[ ] Set $W := \emptyset$;
\item[ ] {\bf while} $V \neq W$ {\bf do}
\item[ ] {\bf begin}
\begin{itemize}
\item[ ] take any maximal vertex of $G$, say $v \in V$, and set $W := W \cup \{v\}$;
\item[ ] compute a minimum (finite) weight e.d.s.\ containing $v$ in the connected component of $G[V]$ with $v$ (if it exists) $\{$by Lemma $\ref{WEDmaxvdirectS123frchordalgr}$ and Procedure $\ref{vMaxWED}\}$;
\item[ ] {\bf if} $Z^-(v) \neq \emptyset$ {\bf then} $\{$by Fact 1$\}$ 
\item[ ] {\bf begin}
\begin{itemize}
\item[ ] {\bf for each} vertex $x \in Z^-(v)$, assign weight $\infty$ to all vertices in $N(v) \setminus N(x)$;
\item[ ] remove $v$ from $V$, i.e., set $V := V \setminus \{v\}$
\end{itemize}
\item[ ] {\bf end}
\end{itemize}
\item[ ] {\bf end}
\end{itemize}

\begin{itemize}
\item[ ] {\bf if} there exist some e.d.s.\ of finite weight of $G$ (in particular, for each resulting set of e.d.s.\ candidates, check whether this is an e.d.s.\ of $G$) {\bf then} choose one of minimum weight, and return it {\bf else} return ``$G$ has no e.d.s.''
\end{itemize}

\noindent
{\bf end}
\end{algo}

\medskip

The correctness and the polynomial time bound of the algorithm is a consequence of the arguments above and in particular of Lemma \ref{WEDmaxvdirectS123frchordalgr} and Fact 1. This completes the proof of Theorem~\ref{WEDdirectS123frchordalgr}.
\qed

\medskip

\begin{figure}[ht]
  \begin{center}
   \epsfig{file=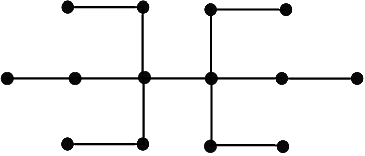}
   \caption{The maximum tree $T$ for which the complexity of ED for $T$-free chordal graphs is open.}
   \label{maxtree-free}
  \end{center}
\end{figure}

It is still an open question how to generalize this approach. For example, the complexity of WED remains an open problem for $S_{2,2,3}$-free chordal as well as for $S_{2,2,2}$-free chordal graphs. However, for trees and forests $T$, there are only finitely many cases for the complexity of WED on $T$-free chordal graphs since
WED on $T$-free chordal graphs is \NP-complete if $T$ contains an induced $K_{1,5}$ or $2P_3$. In Figure \ref{maxtree-free}, the maximum tree without induced $K_{1,5}$ and $2P_3$ is shown.

\section{Conclusion}

The results described in Theorems \ref{EDnetfrchordalG2}, \ref{EDextgemfrchordalG2}, and \ref{WEDdirectS123frchordalgr} are still far away from a dichotomy for the complexity of ED on $H$-free chordal graphs. For chordal graphs $H$ with four vertices, all cases are solvable in polynomial time as described in Lemma \ref{WEDOpenHfrchordalgr} below.

\begin{figure}[ht]
  \begin{center}
   \epsfig{file=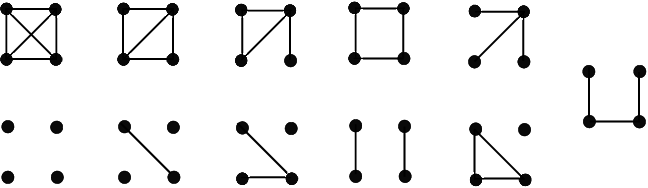}
   \caption{All graphs $H$ with four vertices}
  \label{4vertexgraphs}
  \end{center}
\end{figure}

For chordal graphs $H$ with five vertices, the complexity of ED on $H$-free chordal graphs is still open for the following graphs as described in Lemma \ref{WEDOpenHfrchordalgr}:

\begin{figure}[ht]
  \begin{center}
   \epsfig{file=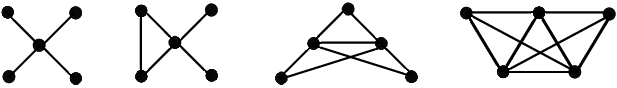}
   \caption{Graphs $H_1,\ldots,H_4$ with five vertices for which ED is open for $H$-free chordal graphs}
   \label{K14dart}
  \end{center}
\end{figure}

\begin{lemma}\label{WEDOpenHfrchordalgr}
\mbox{ }
\begin{itemize}
\item[$(i)$] For every chordal graph $H$ with exactly four vertices, WED is solvable in polynomial time for $H$-free chordal graphs.
\item[$(ii)$] For every chordal graph $H$ with exactly five vertices, the four cases described in Figure \ref{K14dart} are the only ones for which the complexity of WED is open for $H$-free chordal graphs.
\end{itemize}
\end{lemma}

\noindent
{\bf Proof.}
$(i)$: It is well known (see \cite{BraDabHuaPau2015}) that for $H \in \{K_4,K_4-e,paw,P_4\}$, the clique-width is bounded for $H$-free chordal graphs and thus, WED is solvable in polynomial time. By Theorem~\ref{EDextgemfrchordalG2} as well as by Theorem \ref{WEDdirectS123frchordalgr},
WED is solvable in polynomial time for claw-free chordal graphs.

By Lemma \ref{P2cojoin}, WED is solvable in polynomial time for all other graphs $H$ with four vertices (see Figure \ref{4vertexgraphs} for all such graphs; clearly, $C_4$ is excluded).

\medskip

\noindent
$(ii)$: For graphs $H$ with five vertices, let $v$ be one of its vertices. We consider the following cases for $N(v)$ (and clearly exclude the cases when $H$ is not chordal):

\medskip

\noindent
{\bf Case 1.} $|N(v)|=4$ (i.e., $v$ is universal in $H$):

\medskip

Clearly, if $H[N(v)]$ is a $2P_2$ then $H$ is a butterfly and thus, WED is \NP-complete. If $H[N(v)]$ is a $K_4$, or paw, or $P_4$, or $K_3+P_1$, then the clique-width is bounded \cite{BraDabHuaPau2015}; in particular, if $H[N(v)]$ is a paw or $K_3+P_1$ then $H$ is an induced subgraph of $\overline{K_{1,3}+2P_1}$, and according to Theorem 1 of \cite{BraDabHuaPau2015}, the clique-width is bounded. If $H[N(v)]$ is $P_3+P_1$ then it is a special case of
Theorem \ref{EDextgemfrchordalG2}, where it is shown that this case can be solved in polynomial time.
The other cases correspond to graphs $H_1,\ldots,H_4$ of Figure \ref{K14dart} (by Theorem 1 of \cite{BraDabHuaPau2015}, their clique-width is unbounded).

\medskip

\noindent
{\bf Case 2.} $|N(v)|=0$ (i.e., $v$ is isolated in $H$): By Lemma \ref{P2cojoin}, and by Lemma \ref{WEDOpenHfrchordalgr} (i), WED is solvable in polynomial time.

\medskip

In particular, for the same reason, WED is solvable in polynomial time whenever $H$ is not connected (since in that case, at least one connected component of $H$ has at most two vertices). Thus, from now on, we can assume that $H$ is connected.

\medskip

\noindent
{\bf Case 3.} $|N(v)|=3$ (and thus, $|\overline{N(v)}|=1$):

\medskip

If $v$ has exactly one non-neighbor in $K_4$ then $H=H_4$.
If $v$ has exactly one non-neighbor in $K_{1,3}$ with midpoint $w$, namely one of degree 1, then $H[N(w)]=P_3+P_1$ according to Case 1 (a special case of
Theorem \ref{EDextgemfrchordalG2}).

If $v$ has exactly one non-neighbor in a diamond, namely one of degree 2, or exactly one non-neighbor in a paw, namely one of degree 1,
then $H$ is an induced subgraph of $\overline{K_{1,3}+2P_1}$.
Moreover, if $v$ has exactly one non-neighbor in a paw, namely one of degree 2, then $H$ is a gem, and if $v$ has exactly one non-neighbor in $P_4$, namely one of degree 1, then $H$ is a co-chair. If $v$ has exactly one non-neighbor in $P_1+P_3$, namely one of degree 1, then $H$ is a bull. In all these cases, the clique-width is bounded according to Theorem 1 of \cite{BraDabHuaPau2015}.

In the remaining cases, $H$ is a chair or co-$P$, and thus, WED is solvable in polynomial time.

\medskip

\noindent
{\bf Case 4.} $|N(v)|=2$ (and thus, $|\overline{N(v)}|=2$):

\medskip

In one of the cases, namely if $v$ is adjacent to the two vertices with degree 1 and with degree~3 in a paw, $H$ is a butterfly and thus, WED is \NP-complete.

If $v$ has exactly two neighbors in $K_4$ or if $v$ is adjacent to degree 2 and degree 3 vertices in diamond or if $v$ is adjacent to the two degree 2 vertices in a paw or if $v$ is adjacent to the two degree 2 vertices (midpoints) in a $P_4$, then by Theorem 1 of \cite{BraDabHuaPau2015}, the clique-width is bounded.

If $v$ is adjacent to the two vertices of degree 3 of a diamond then $H=H_3$.
If $v$ is adjacent to degree 2 vertex $u$ and degree 3 vertex $w$ in a paw then for the degree 3 vertex $w$, $H[N(w)]=P_3+P_1$ as above.
If $v$ is adjacent to degree 1 and degree 3 vertices in a claw then $H=H_2$.

In all other cases, $H$ is a $P_5$, chair or co-$P$, and thus, WED is solvable in polynomial time (by Theorem \ref{EDextgemfrchordalG2} for co-$P$-free chordal graphs, and by Theorems \ref{EDextgemfrchordalG2} and \ref{WEDdirectS123frchordalgr}, for $P_5$-free chordal graphs, and for chair-free chordal graphs).

\medskip

\noindent
{\bf Case 5.} $|N(v)|=1$ (and thus, $|\overline{N(v)}|=3$):

\medskip

Now $v$ is adjacent to exactly one vertex of $V \setminus \{v\}$.

If $v$ is adjacent to a degree 3 vertex $w$ of a diamond then $H[N(w)]=P_3+P_1$ as above.
If $v$ is adjacent to a degree 3 vertex of a paw then $H=H_2$.
If $v$ is adjacent to a degree 3 vertex of a claw then $H=H_1$.

If $v$ is adjacent to one vertex of $K_4$ or one vertex of the diamond of degree 2 (co-chair) or one vertex of a paw of degree 2 (bull) then by Theorem 1 of \cite{BraDabHuaPau2015}, the clique-width is bounded.

In all other cases, $H$ is a $P_5$, chair or co-$P$, and thus, WED is solvable in polynomial time as above.
\qed

\medskip

Of course there are many larger examples of graphs $H$ for which ED is open for $H$-free chordal graphs. In general, one can restrict $H$ by various conditions such as diameter (if the diameter of $H$ is at least 6 then $H$ contains an induced $2P_3$) and size of connected components (if $H$ has at least two connected components of size at least 3 then $H$ contains an induced $2P_3$, $K_3+P_3$, or $2K_3$). It would be nice to classify the open cases in a more detailed way.

\medskip

\noindent
{\bf Acknowledgment.} We gratefully thank the anonymous reviewers for their comments and corrections.
The second author would like to witness that he just tries to pray a lot and is not able to do anything without that - ad laudem Domini.

\begin{footnotesize}

\end{footnotesize}


\begin{thebibliography}{99}

\bibitem{BanBarSla1988}
    D.W.~Bange, A.E.~Barkauskas, and P.J.~Slater,
    Efficient dominating sets in graphs,
    in: R.D. Ringeisen and F.S. Roberts, eds., Applications of Discrete Math. (SIAM, Philadelphia, 1988) 189-199.

\bibitem{BanBarHosSla1996}
    D.W.~Bange, A.E.~Barkauskas, L.H.~Host, and P.J.~Slater,
    Generalized domination and efficient domination in graphs,
    {\sl Discrete Math.} 159 (1996) 1-11.

\bibitem{Biggs1973}
   N.~Biggs,
   Perfect codes in graphs,
   {\sl J. of Combinatorial Theory (B)}, 15 (1973) 289-296.

\bibitem{BraDabHuaPau2015}
    A.~Brandst\"adt, K.K.~Dabrowski, S.~Huang, and D.~Paulusma,
   Bounding the clique-width of $H$-free chordal graphs,
   {\sl J. of Graph Theory} 86 (2017) 42-77.

\bibitem{BraEscFri2015}
    A.~Brandst\"adt, E.~Eschen, and E.~Friese,
    Efficient domination for some subclasses of $P_6$-free graphs in polynomial time,
    extended abstract in: Proceedings of WG 2015, E.W.~Mayr, ed., LNCS 9224, pp. 78-89, 2015; full version in: CoRR arXiv:1503.00091, 2015.

\bibitem{BraEscFriKar2017}
    A.~Brandst\"adt, E.~Eschen, E.~Friese, and T.~Karthick,
    Efficient domination for classes of $P_6$-free graphs,
    {\sl Discrete Applied Math.} 223 (2017) 15-27.

\bibitem{BraFicLeiMil2013}
  A.~Brandst{\"a}dt, P.~Fi{\v c}ur, A.~Leitert, and M.~Milani{\v c},
  Polynomial-time algorithms for Weighted Efficient Domination problems in AT-free graphs and dually chordal graphs,
 {\sl Information Processing Letters} 115 (2015) 256-262.

\bibitem{BraGia2014}
    A.~Brandst\"adt and V.~Giakoumakis,
    Weighted Efficient Domination for $(P_5+kP_2)$-Free Graphs in Polynomial Time,
    CoRR arXiv:1407.4593, 2014.

\bibitem{BraGiaMil2018}
    A.~Brandst\"adt, V.~Giakoumakis, and M.~Milani\v c,
    Weighted efficient domination for some classes of $H$-free and of $(H_1, H_2)$-free graphs,
    {\sl Discrete Applied Math.} 250 (2018) 130-144.

\bibitem{BraLeiRau2012}
    A.~Brandst\"adt, A.~Leitert, and D.~Rautenbach,
    Efficient Dominating and Edge Dominating Sets for Graphs and Hypergraphs,
    extended abstract in: Conference Proceedings of ISAAC 2012, LNCS 7676, 2012, 267-277.

\bibitem{BraLeSpi1999}
    A.~Brandst\"adt, V.B.~Le, and J.P.~Spinrad,
    Graph Classes: A Survey,
    {\sl SIAM Monographs on Discrete Math. Appl., Vol.} 3,
    SIAM, Philadelphia (1999).

\bibitem{BraMilNev2013}
    A.~Brandst\"adt, M.~Milani\v c, and R.~Nevries,
    New polynomial cases of the weighted efficient domination problem,
    extended abstract in: Conference Proceedings of MFCS 2013, LNCS 8087, 195-206.
    Full version: CoRR arXiv:1304.6255, 2013.

\bibitem{BraMos2015}
    A.~Brandst\"adt and R.~Mosca,
    Weighted efficient domination for $P_6$-free graphs in polynomial time,
    CoRR arXiv:1508.07733, 2015

\bibitem{BraMos2016}
    A.~Brandst\"adt and R.~Mosca,
    Weighted efficient domination for $P_5$-free and $P_6$-free graphs,
    extended abstract in: Proceedings of WG 2016, P. Heggernes, ed., LNCS 9941, pp. 38-49, 2016. Full version:
    {\sl SIAM J. Discrete Math.} 30, 4 (2016) 2288-2303.

\bibitem{ChaLiu1993}
M.-S.~Chang and Y.C.~Liu,
    Polynomial algorithms for the weighted perfect domination problems on chordal graphs and split graphs,
    {\sl Information Processing Letters} 48 (1993) 205-210.

\bibitem{ChaPanCoo1995}
    G.J.~Chang, C.~Pandu Rangan, and S.R.~Coorg,
    Weighted independent perfect domination on co-comparability graphs,
    {\sl Discrete Applied Math.} 63 (1995) 215-222.

\bibitem{CouMakRot2000}
   B.~Courcelle, J.A.~Makowsky, and U.~Rotics,
   Linear time solvable optimization problems on graphs of bounded clique-width,
   {\sl Theory of Computing Systems} {\bf 33} (2000) 125-150.

\bibitem{EscWan2014}
  E.M.\ Eschen and X.\ Wang,
  Algorithms for unipolar and generalized split graphs,
  {\sl Discrete Applied Mathematics} 162 (2014) 195-201.

\bibitem{FoeHam1977}
S.~F\H{o}ldes and P.L.~Hammer,
    Split graphs,
    {\sl Congressus Numerantium} 19 (1977) 311-315.

\bibitem{Frank1975}
    A.~Frank,
    Some polynomial algorithms for certain graphs and hypergraphs,
    Proceedings of the 5th British Combinatorial Conf. (Aberdeen 1975),
    {\sl Congressus Numerantium} XV (1976) 211-226.

\bibitem{GarJoh1979}
    M.R.~Garey and D.S.~Johnson,
    Computers and Intractability--A Guide to the Theory of NP-completeness,
    {\sl Freeman}, San Francisco, 1979.

\bibitem{GolRot2000}
    M.C.~Golumbic and U.~Rotics,
    On the Clique-Width of Some Perfect Graph Classes,
    {\sl Internat. J. Foundations of Computer Science} 11 (2000) 423-443.

\bibitem{Karp1972}
    R.M.~Karp,
    Reducibility among combinatorial problems,
    In: {\em Complexity of Computer Computations}, Plenum Press, New York (1972) 85-103.

\bibitem{LiaLuTan1997}
   Y.D.~Liang, C.L.~Lu, and C.Y.~Tang,
   Efficient domination on permutation graphs and trapezoid graphs,
   in: {\sl Proceedings COCOON'97}, T. Jiang and D.T. Lee, eds., {\sl Lecture Notes in Computer Science} Vol. 1276
   (1997) 232-241.

\bibitem{Lin1998}
   Y.-L.~Lin,
   Fast algorithms for independent domination and efficient domination in trapezoid graphs,
   in: {\sl Proceedings ISAAC'98}, {\sl Lecture Notes in Computer Science} Vol. 1533 (1998) 267-275.

\bibitem{LivSto1988}
   M.~Livingston and Q.~Stout,
   Distributing resources in hypercube computers,
   in: {\sl Proceedings 3rd Conf. on Hypercube Concurrent Computers and Applications} (1988) 222-231.

\bibitem{LokPilvan2015}
    D.~Lokshtanov, M.~Pilipczuk, and E.J.~van~Leeuwen,
    Independence and Efficient Domination on $P_6$-Free Graphs,
    Proceedings of the ACM-SIAM Symposium on Discrete Algorithms (SODA) 2016, 1784-1803.

\bibitem{LuTan1998}
    C.L.~Lu and C.Y.~Tang,
    Solving the weighted efficient edge domination problem on bipartite permutation graphs,
    {\sl Discrete Applied Math.} 87 (1998) 203-211.

\bibitem{LuTan2002}
    C.L.~Lu and C.Y.~Tang,
    Weighted efficient domination problem on some perfect graphs,
    {\sl Discrete Applied Math.} 117 (2002) 163-182.

\bibitem{Milan2012}
    M.~Milani\v c,
    Hereditary Efficiently Dominatable Graphs,
    {\sl Journal of Graph Theory} 73 (2013) 400-424.

\bibitem{SmaSla1995}
C.B.~Smart and P.J.~Slater,
Complexity results for closed neighborhood order parameters,
{\sl Congr. Numer.} 112 (1995) 83-96.

\bibitem{Yen1992}
    C.-C.~Yen,
    Algorithmic aspects of perfect domination,
    Ph.D. Thesis, Institute of Information Science, National Tsing Hua University, Taiwan 1992.

\bibitem{YenLee1996}
    C.-C.~Yen and R.C.T.~Lee,
    The weighted perfect domination problem and its variants,
{\sl Discrete Applied Math.} 66 (1996) 147-160.

\bibitem{Zvero2006}
  I.E. Zverovich, Satgraphs and independent domination. Part 1,
  {\sl Theoretical Computer Science} 352 (2006) 47-56.

\end{thebibliography}
\end{document}